\documentclass[%
superscriptaddress,
 amsmath,amssymb,
 aps,
nofootinbib,
]{revtex4-2}

\usepackage{layouts}
\usepackage{comment}
\usepackage{tabularx}
\usepackage{graphicx}
\usepackage{dcolumn}
\usepackage{siunitx}
\usepackage{bm}
\usepackage{mathtools}
\usepackage{color}
\newcommand{\ket}{\rangle}
\newcommand{\bra}{\langle}

\newcommand{\degree}{$^{\circ}$}

\newcolumntype{x}[1]{>{\centering\let\newline\\\arraybackslash\hspace{0pt}}p{#1}}

\newcommand{\sixJ}[6] {\left\{\begin{array}{ccc} #1 & #2 & #3 \\ #4 & #5 & #6 \end{array}\right \}}
\newcommand{\threeJ}[6] {\left(\begin{array}{ccc} #1 & #3 & #5 \\ #2 & #4 & #6 \end{array}\right )}

\newcommand{\X}[0] {$\tilde{X}(000)$}
\newcommand{\Xbend}[0] {$\tilde{X}(010)$}
\newcommand{\A}[0] {$\tilde{A}(000)$}

\newcommand{\XSigbend}[0] {$\tilde{X} {}^2\Sigma^+ (010)$}
\newcommand{\APi}[0] {$\tilde{A}{}^2\Pi_{1/2}(000)$}

\usepackage{rotating}

\usepackage[dvipsnames]{xcolor}

\usepackage{threeparttable}

\usepackage{here}
\selectfont

\begin{document}

\title{Characterizing the Fundamental Bending Vibration of a Linear Polyatomic Molecule for Symmetry Violation Searches}

\author{Arian Jadbabaie}
\thanks{These authors contributed equally.}
\affiliation{California Institute of Technology, Division of Physics, Mathematics, and Astronomy.  Pasadena, CA 91125}

\author{Yuiki Takahashi}
\thanks{These authors contributed equally.}
\affiliation{California Institute of Technology, Division of Physics, Mathematics, and Astronomy.  Pasadena, CA 91125}

\author{Nickolas H. Pilgram}
\altaffiliation[Current affiliation: ]{NIST Physical Measurement Laboratory.  Gaithersburg, MD 20899}
\affiliation{California Institute of Technology, Division of Engineering and Applied Science.  Pasadena, CA 91125}

\author{Chandler J. Conn}
\affiliation{California Institute of Technology, Division of Physics, Mathematics, and Astronomy.  Pasadena, CA 91125}

\author{Yi Zeng}
\affiliation{California Institute of Technology, Division of Physics, Mathematics, and Astronomy.  Pasadena, CA 91125}

\author{Chi Zhang}
\affiliation{California Institute of Technology, Division of Physics, Mathematics, and Astronomy.  Pasadena, CA 91125}

\author{Nicholas R. Hutzler}
\affiliation{California Institute of Technology, Division of Physics, Mathematics, and Astronomy.  Pasadena, CA 91125}

\date{\today}

\begin{abstract}

Polyatomic molecules have been identified as sensitive probes of charge-parity violating and parity violating physics beyond the Standard Model (BSM). For example, many linear triatomic molecules are both laser-coolable and have parity doublets in the ground electronic $\tilde{X} {}^2\Sigma^+  (010)$ state arising from the bending vibration, both features that can greatly aid BSM searches. Understanding the $\tilde{X} {}^2\Sigma^+ (010)$ state is a crucial prerequisite to precision measurements with linear polyatomic molecules. Here, we characterize fundamental bending vibration of ${}^{174}$YbOH using high-resolution optical spectroscopy on the nominally forbidden \XSigbend{}${}\rightarrow{}$\APi{} transition at 588~nm. We assign 39 transitions originating from the lowest rotational levels of the \XSigbend{} state, and accurately model the state's structure with an effective Hamiltonian using best-fit parameters. Additionally, we perform Stark and Zeeman spectroscopy on the \XSigbend{} state and fit the molecule-frame dipole moment to $D_\mathrm{mol}=2.16(1)$ D and the effective electron \textit{g}-factor to $g_S=2.07(2)$. Further, we use an empirical model to explain observed anomalous line intensities in terms of interference from spin-orbit and vibronic perturbations in the excited \APi{} state.  Our work is an essential step toward searches for BSM physics in YbOH and other linear polyatomic molecules.

\end{abstract}

\maketitle

\section{Introduction}

Polyatomic molecules are at the frontier of advanced control over quantum complexity. Their additional rovibrational degrees of freedom provide a large degree of control and tunability of both molecular structure and interactions with a wide range of applications. Rapid progress~\cite{Isaev2016Poly,Kozyryev2016Poly,Fitch2021Review} has been made in laser cooling molecules, including polyatomic CaOH~\cite{Baum20201dmot,Baum2021cycling}, CaOCH$_3$~\cite{Mitra2020}, SrOH~\cite{Kozyryev2016SrOH,Kozyryev2017SrOH}, and YbOH~\cite{Augenbraun2020}.  Recently, CaOH was optically trapped and laser-cooled to ultracold temperatures~\cite{Vilas2022MOT,Hallas2022ODT}. Quantum control of polyatomic molecules will benefit next-generation searches for new physics beyond the Standard Model~\cite{Kozyryev2017PolyEDM, Kozyryev2021DM, Yu2021STM,Hutzler2020Review}, and will enable advances in quantum computation, simulation, and chemistry \cite{Yu2019STM,Albert2020,Wall2015,Yang2022}.

Currently, measurements of diatomic ThO and HfF$^+$ bound charge-parity (CP) violating new physics at TeV energy scales~\cite{ACME2018, Cairncross2017}. These experiments benefit significantly from parity doubling, the occurrence of nearly-degenerate levels of opposite parity. Molecules with parity doublets can be easily aligned in the lab frame with the application of modest electric fields. Furthermore, when polarized, these molecules have both aligned and anti-aligned states. Known as internal co-magnetometers, these states allow for reversal of CP-violating interactions without modifying the external lab field~\cite{Eckel2013}.  This degree of control over molecular alignment is highly advantageous for robust systematic error rejection in searches for CP violation. In diatomic molecules, parity doublets require orbital angular momentum, which conflicts with electronic requirements for efficient laser cooling, especially for heavy molecules with enhanced sensitivity to new physics~\cite{Kozyryev2017PolyEDM,Hutzler2020Review}. 

Polyatomic molecules offer both generic parity doublets and laser cooling, and therefore provide a route to significantly improve constraints on new CP-violating physics by multiple orders of magnitude~\cite{Kozyryev2017PolyEDM}. A number of CP violation searches are underway with laser-coolable diatomic molecules, such as BaF~\cite{Aggarwal2018}, YbF~\cite{Lim2018,Fitch2020}, TlF~\cite{Grasdijk2021TlF}, and RaF~\cite{Garcia2020RaF,Petrov2020RaF}. Without parity doublets in their ground states, these molecules require large electric fields ($>$10 kV/cm) for significant polarization. By contrast, molecules with parity doublets offer similar polarization in much smaller fields, and the variety of molecular orientations offer richer possibilities for state tuning. In polyatomic molecules, parity doublets arise from rotation around the inter-nuclear axis and exist independently of the electronic structure used for laser cooling~\cite{Isaev2016Poly,Kozyryev2016Poly,Kozyryev2017PolyEDM}. Examples of polyatomic parity doublets include $K$ doublets in rotations of symmetric molecules, asymmetry doublets in the rotations of asymmetric molecules, and $\ell$ doublets in bending modes of linear polyatomic molecules. 

YbOH molecules in their doubly-degenerate bending mode have been identified as sensitive probes of CP-violating physics~\cite{Kozyryev2017PolyEDM}. The Yb-centered, core-penetrating valence electron provides both new physics sensitivity and optical cycling, which was demonstrated with Sisyphus cooling of a YbOH beam to a transverse temperature of $<$600~$\si{\micro\kelvin}$~\cite{Augenbraun2020}. Meanwhile, the vibrational bending motion provides $\ell$-type parity doublets that allow polarization control and internal co-magnetometry in modest external fields. Furthermore, the multiple stable isotopes of Yb provide opportunities for CP violation searches in both the hadronic and leptonic sectors of the Standard Model~\cite{Flambaum2014,Kozyryev2017PolyEDM,Maison2019,Prasannaa2019,Denis2019,Gaul2020,Zakharova2021YbOH,Petrov2022}.  Finally, other experiments leveraging the bending motion of linear triatomic molecules, including CP violation searches with SrOH~\cite{Lasner2022} and RaOH~\cite{Isaev2017RaOH,Kozyryev2017PolyEDM}, and parity-violation searches with linear triatomics~\cite{Norrgard2019}, warrant further investigation of these states, for which there is no previous, complete study of all molecular properties.

Here, we present a high-resolution, optical spectroscopy study of the fundamental bending vibration in the electronic ground state of ${}^{174}$YbOH. The spectra are obtained by laser excitation on a rovibrationally forbidden electronic transition in a cryogenic buffer gas beam (CBGB).  By analyzing the field-free, Stark, and Zeeman spectra, we model the rotational structure of the bending molecule, characterize the electric and magnetic tuning of the levels, and extract the molecule-frame dipole moment. Our results demonstrate the high level of control available in polyatomic molecules, which will be useful for future symmetry violation searches. 

The structure of the paper is as follows. First, we provide a brief overview of the overall molecular structure in section \ref{sec:notation}. The methods are described in section \ref{sec:methods}, with section \ref{sec:expt} describing the experimental apparatus, and section \ref{sec:theory} describing the effective Hamiltonians used to model the molecular states. In section \ref{sec:results} we describe our experimental results and analysis. Section \ref{sec:lines} discusses the field-free spectrum and optimal state parameters, section \ref{sec:intensity} describes our model for the anomalous line intensities of the forbidden transition, and section \ref{sec:field} presents the Stark and Zeeman spectra and their analysis. We conclude in section \ref{sec:conclusion}.

\subsection{Molecular Structure}\label{sec:notation}

In this section, we briefly review the structure of linear polyatomic molecules, including states with bending vibration.  We label the ground and excited state electronic states as $\tilde{X}$ and $\tilde{A}$, respectively. Electronic states of linear polyatomic molecules are labeled with the term symbol ${}^{2S+1} \Lambda_{\Omega} (v_1 \, v^{l}_2 \, v_3)$, where $\Lambda = \vec{L}\cdot \hat{n}$ is the projection of electronic orbital angular momentum $L$ on the internuclear axis $\hat{n}$, $\Sigma=\vec{S}\cdot \hat{n}$ is the projection of the electron spin $S$, $\Omega = \Lambda + \Sigma=\vec{J}\cdot\hat{n}$ is the total projection of the spin and rotational angular momentum $J$, and $v_i$ denotes the number of quanta in the three vibrational modes of the molecule. For $\Lambda=0$ states, an additional $+/-$ subscript is used to denote the parity of the electronic configuration, and the $\Omega$ subscript is sometimes dropped. In YbOH~\cite{Kozyryev2017PolyEDM}, the $v_1$ mode is the Yb-O stretch, the $v_3$ mode the O-H stretch, and, due to the Yb mass, the doubly-degenerate $v^{\ell}_2$ mode can be viewed as the bending of the H atom relative to the Yb-O axis~\cite{Li1995}. The additional $\ell$ label denotes the number of quanta of vibrational angular momentum $G$ projected on the internuclear axis, $\ell=\vec{G}\cdot \hat{n}$. The degeneracy of $\pm \ell$ states are lifted by higher order perturbations, giving rise to parity doublets~\cite{Herzberg1942,Nielsen1951}.

The above electronic labeling scheme treats the vibrational degrees of freedom separately. However, for states with non-zero $\ell$ and $\Lambda$, interactions of the electrons with the bending vibration, known as Renner-Teller couplings~\cite{Herzberg1967,Brown2000}, will cause rovibrational splittings for different states of $K=\Lambda + \ell = \vec{N}\cdot \hat{n}$. Here is $\vec{N} = \vec{J} - \vec{S}$ is the rovibrational angular momentum of the electrons and nuclei, excluding spin. Note that $N$ can receive contributions from multiple sources: the end-over-end molecular rotation $R$, electronic orbital angular momentum $L$, and vibrational angular momentum $G$.  When both Renner-Teller and spin-orbit couplings are present, neither $K$ nor $\Omega$ are completely conserved, and instead the eigenstates have well defined projection quantum number $P = \vec{J}\cdot \hat{n} = \Lambda + \ell + \Sigma$. We note that the total angular momentum cannot be less than the projection angular momentum. For example, in a state with well-defined $N$ and $K$, we always have $N\geq |K|$; a consequence relevant for this work is that the lowest rotational level of an $\ell=1$ bending mode has $N=1$. 

We will restrict our discussion to states with $v_1 = v_3 = 0$ and $v_2 \in \{1,0\}$, allowing us to write vibronic term symbols as ${}^{2S+1} K_{P}$.  Note that in the term symbols, both $\Lambda$ and $K$ are denoted as $\Sigma,\Pi,\Delta,\ldots$ to indicate 0, 1, 2, $\ldots$, similar to the $S, P, D, \ldots$ notation in atoms.  This can lead to confusion; for example the (010) vibrational state in the ground electronic state is a $^2\Sigma^+$ electronic state, but a $^2\Pi$ vibronic state. Whenever we do not include the $(v_1\,v_2\,v_3)$ label, we are referring to a vibronic term, unless otherwise noted.

In this work, we study the $\tilde{X}{}^2\Sigma^+_{1/2}(01^10) \rightarrow \tilde{A} {}^2\Pi_{1/2}(000)$ band of ${}^{174}$YbOH. This transition is nominally forbidden in the dipole approximation, which requires $\Delta \ell=0$, and it occurs via intensity borrowing in the excited state, as we discuss later. We will neglect the other spin-orbit manifold, $\tilde{A}{}^2\Pi_{3/2}(000)$, which is located $\sim$40 THz above $\tilde{A}{}^2\Pi_{1/2}(000)$. The large spin-orbit coupling in YbOH means $\Omega$ is an approximately good quantum number, even in bending states. For simplicity, we will abbreviate the ground state label as \Xbend{} and the excited state label as \A{}. 

In ${}^{174}$YbOH, the ${}^{174}$Yb nucleus has no nuclear spin, and the hyperfine structure from the distant hydrogen nuclear spin $I$ is optically unresolved~\cite{Nakhate2019} and only contributes to broadening in the ground state. Therefore in this study we neglect $I$, and label states with well-defined total angular momentum $J$.

Ground state quantum numbers are denoted with a double prime, e.g. $N^{\prime\prime}$, and excited states with a single prime, e.g. $J^{\prime}$. We denote rotational lines with notation similar to Ref.~\cite{Steimle2019}. Given the parity doubling in both \Xbend{} and \A{}, we add an additional label to denote the parity of the ground state. We label transitions as $^{\Delta N} \Delta J^{\mathcal{P}^{\prime\prime}}_{F^{\prime}_i, F^{\prime\prime}_i} (N^{\prime\prime})$. Here, $F^{\prime}_i = 1$ for the excited state, $F^{\prime\prime}_i= 1, 2$ denotes ground states with $J^{\prime\prime} = N^{\prime\prime} \pm S$, and $\mathcal{P}^{\prime\prime}=\pm$ denotes the ground state parity.

\section{Methods}\label{sec:methods}

\subsection{Experiment: Apparatus and Signals}\label{sec:expt}

\begin{figure}[t]
    \centering
    \includegraphics[width=0.6\textwidth]{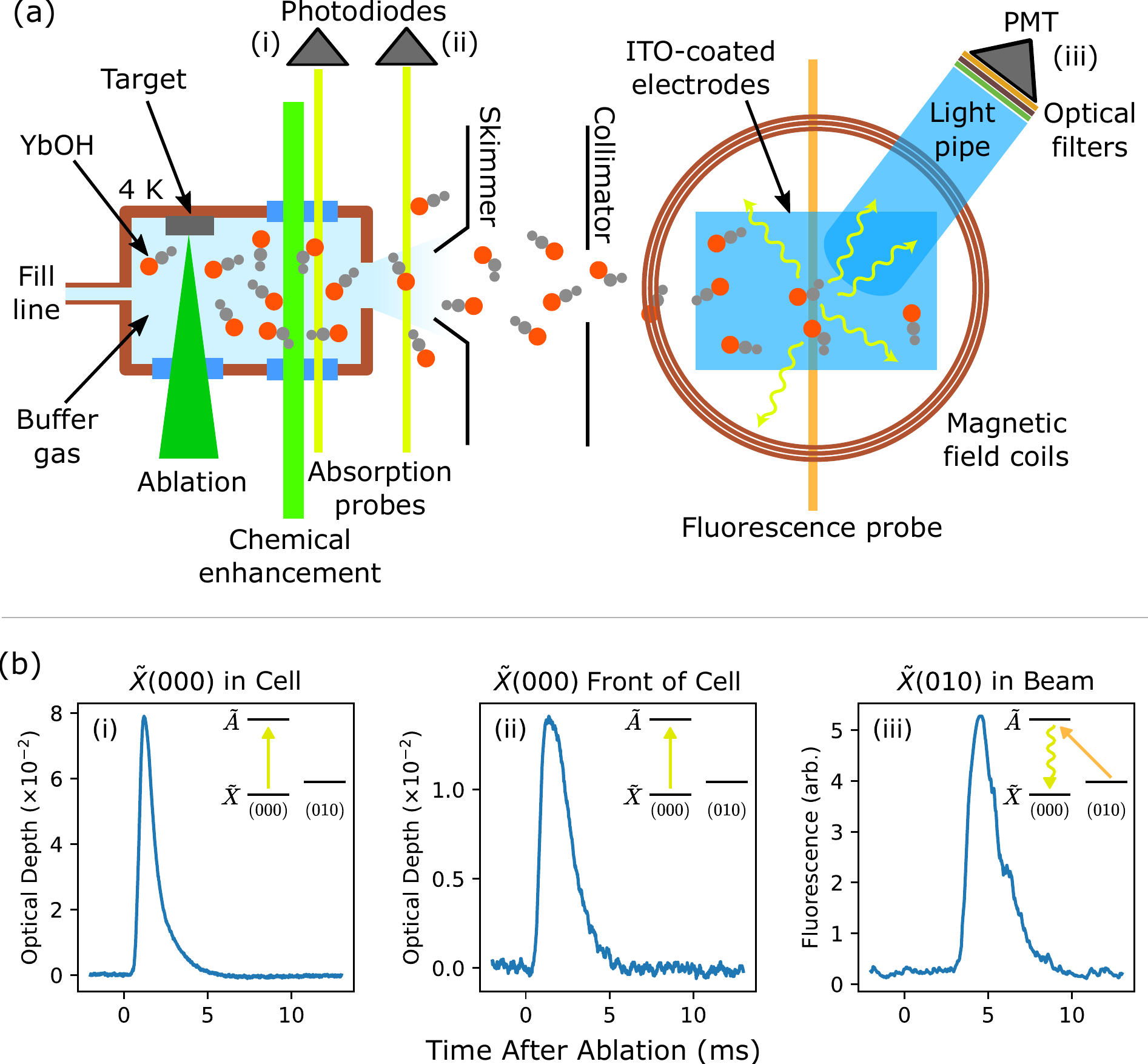}
    \caption{\label{fig:experiment} (a) Experimental schematic. YbOH molecules are produced in the 4 K cryogenic buffer gas cell (brown box) by laser ablation (dark green triangle) of a solid pressed target. The molecules are thermalized by collisions with He buffer gas continuously flowed into the cell. Chemical production of YbOH is enhanced by exciting Yb atoms using a laser (light green line) resonant with the ${}^1 S_0 \rightarrow {}^3 P_1$ atomic Yb transition. Some of the molecules are produced in the \Xbend{} bending mode. The molecules are entrained in the He gas flow and extracted out of the cell. We detect the molecule number density in the \X{} state via absorption spectroscopy (yellow lines) both in the cell (i) and in front of the cell (ii). The molecular beam is collimated by a skimmer and collimators before entering the probe region with electric and magnetic fields. We apply magnetic fields using coils outside the vacuum chamber, and apply electric fields using ITO coated glass electrodes inside the vacuum chamber. In the center of the fields, molecules in the \Xbend{} state are excited by a laser (orange line) and their fluorescence is collected through a light pipe to a PMT (iii).
    (b) Sample signals from the CBGB. (i) In-cell absorption on the $^R R_{11}(0)$ line of YbOH $\tilde{X}(000) \rightarrow \tilde{A}(000)$. The peak optical depth corresponds to a molecule density of $\sim$5$\times 10^{9}$~cm$^{-3}$ in the $\tilde{X}(000)$, $N=0$ state.  (ii) Front of cell absorption on the same $^R R_{11}(0)$ line. The peak optical depth corresponds to a molecule density of $\sim$2$\times10^9$~cm$^{-3}$. (iii) Fluorescence after excitation of the bending mode on a strong $\tilde{X}(010) \rightarrow \tilde{A}(000)$ line. The integrated signal corresponds to $\sim$8300 photons detected on the PMT.}
\end{figure}

The cryogenic buffer gas beam (CBGB) apparatus (Fig.~\ref{fig:experiment}a) is similar to that from our previous work~\cite{Jadbabaie2020, Pilgram2021}. In summary, the buffer gas cell is formed from a copper block with an interior cylindrical bore 7.5~cm long and 12.7~mm in diameter, with windows on the sides for optical access. The cell is surrounded by radiation shields and cooled by a pulse tube refrigerator down to $\sim$4 K. Helium buffer gas is introduced in the back of the cell via a 3.2~mm gas inlet tube, and passes through a diffuser 3.2~mm downstream in the cell. Typical flow rates are $3-6$ standard cubic centimeters per minute (SCCM). The buffer gas exits the cell via a 5~mm diameter aperture at the front of the cell. Activated charcoal fins on the interior surface of the 4 K radiation shields provide efficient cryo-pumping of the He buffer gas. 

YbOH molecules are produced by laser ablation of pressed powder targets made from a 1:1 stoichiometric mixture of Yb(OH)$_3$ powder and Yb powder (see supplementary materials). Laser ablation is performed by a Nd:YAG laser at 532~nm with $\sim$10 ns pulse length, $25-40$~mJ pulse energy, and $\sim$9 Hz repetition rate. The ablation laser is focused with a 300 or 400~mm lens placed approximately one focal length away from the target. Hot molecules produced via ablation are subsequently thermalized by collisions with $\sim$4~K He buffer gas atoms~\cite{Hutzler2012}. We further increase YbOH yield by around an order of magnitude by exciting atomic Yb to the excited ${}^3P_1$ state~\cite{Jadbabaie2020}. Specifically, we send $\sim$300~mW of 556~nm light into the cell to resonantly drive the ${}^1S_0 \rightarrow {}^3P_1$ transition of ${}^{174}$Yb. This technique significantly increases the quantity of YbOH in excited vibrational states, including the \Xbend{} state, whose population is increased by a factor of $\sim$10.  

A few milliseconds after ablation, the He gas flow extracts the molecules out of the cell through the aperture. Molecule density is monitored both in the cell and outside the cell aperture with 577~nm absorption probes resonant with the $^RR_{11}(0)$ line of the $\tilde{X}(000) \rightarrow \tilde{A}(000)$ transition at 17325.0365~cm$^{-1}$~\cite{Steimle2019}. The extracted beam is rotationally and translationally cold, but can have significant excited vibrational population, a result of inefficient vibrational thermalization from buffer gas collisions~\cite{Kozyryev2015}. This provides a significant advantage, as we obtain $\sim$10$^9$ molecules exiting the cell in the excited bending mode as a result. The molecular beam is collimated by a 6.4~mm diameter skimmer 4.8~cm downstream from the cell aperture, a 9.5~mm diameter hole 11.4~cm downstream from the cell aperture, and a 5~mm diameter hole 23.7~cm downstream from the cell aperture. The beam travels at $150- 200$~m/s toward the laser-induced fluorescence (LIF) measurement region located $\sim$60~cm downstream from the cell. The region is pumped by multiple turbomolecular pumps, and typical pressures when flowing He gas are $1-5 \times 10^{-7}$~Torr. 

In YbOH, the $\tilde{A} (000) \rightarrow \tilde{X} (010)$ transition has a vibrational branching ratio of $r_{010} = 0.054(4)\%$~\cite{Zhang2021VBR}, and the lifetime of the $\tilde{A} {}^2 \Pi_{1/2}$ state is $\tau = 20(2)~\mathrm{ns}$~\cite{Mengesha2020YbOH}. The excited state population primarily decays to the vibrational ground state, $\tilde{X} (000)$, with $r_{000} = 89.44\%$ branching. Therefore, in our experiment, the fluorescence signal will saturate after roughly one photon scatter as the molecules are optically pumped out of the bending mode and mostly into the ground state. With a $\sim$1~mm Gaussian laser beam intersecting a $\sim$200~m/s molecular beam, we can estimate the saturation parameter required for a single photon scatter as $s\approx 1\times10^{-2}$. Using the definition of saturation intensity for a transition with branching ratio $r$ as $I_s = \pi h c/(\lambda^3 \tau r)$~\cite{Wall2008}, we compute an intensity of $I \approx 280$~mW/cm$^2$ required to optically pump the forbidden transition $\tilde{X} (010) \rightarrow \tilde{A} (000)$. For a 1~mm diameter Gaussian laser beam, this requires $\gtrsim\,$2~mW of optical power. While we have neglected rotational branching and other experimental imperfections in this analysis, we observe the power requirements needed to produce fluorescence on such a forbidden line are feasible. 

Downstream in the LIF region, molecules in the $\tilde{X}(010)$ bending mode are excited by a 588~nm laser resonant with the nominally forbidden $\tilde{X}(010)\rightarrow \tilde{A}(000)$ transition. The laser beam, with a $\sim$1~mm diameter and $\sim$40~mW of power, is sent perpendicular to the molecular beam (see Fig \ref{fig:experiment}a) through windows at Brewster's angle. The resulting 577~nm fluorescence from decays to the $\tilde{X}(000)$ state is collected with a 19.4~mm diameter fused-quartz light pipe. A 25.4~mm diameter, 19~mm focal length retroreflecting concave mirror opposite the light pipe improves collection efficiency. We filter out the 588~nm scattered background light using a combination of interference and colored glass filters on the exit of the light pipe, obtaining a signal-to-noise ratio of $>$10. The fluorescence signal is incident on a photomultiplier tube (PMT) module (Hamamatsu H13543-300), and the resulting photocurrent is amplified with a $10^{-8}$ A/V trans-impedance amplifier with a 1.5 kHz low pass filter. 

To obtain the field-free spectrum, we scan the 588~nm probe laser and record its frequency using a wavelength meter (HighFinesse WS7-30) with an absolute accuracy of 30~MHz and a measurement resolution of 1~MHz. To improve the absolute accuracy, we use the probe light to co-record sub-Doppler I$_2$ spectra, obtained with amplitude modulated saturated absorption spectroscopy~\cite{Salumbides2006}. Calibration of the laser frequency using the I$_2$ spectra results in one standard deviation error of 2.35~MHz in absolute frequency accuracy.

Figure \ref{fig:experiment}b shows typical absorption and LIF signals obtained in a single shot. The LIF signal size typically varies from shot to shot due to ablation yield fluctuations. To construct the field-free spectrum, we scan the laser at approximately 1-2~MHz per shot, average the LIF signal for 4 shots, integrate over the molecule pulse duration, and plot the data against the calibrated probe frequency. The observed peaks are fit well by a Lorentzian function, with fitting errors $<3$~MHz. For the Stark and Zeeman spectra, we step the laser in 3~MHz increments, and average the LIF signal for 10 shots at each step. 

For Stark spectroscopy, we use two indium tin oxide (ITO) coated glass plates separated by a 4.99(3)~mm gap to apply fields up to 265~V/cm in the LIF region. Before entering the field region, the molecular beam is further collimated with a 3~mm hole in a grounded aluminum plate. The molecules traveling through the ITO plates are then excited by the 588~nm laser (see Fig.~\ref{fig:experiment}a). The resulting fluorescence is collected through the glass plates with the setup described earlier. For Zeeman spectroscopy, we generate magnetic fields of $0 - 70$ Gauss using two pairs of wire coils outside the vacuum chamber (see Fig.~\ref{fig:experiment}a). The two coil pairs have a diameter of 21.4~cm with 500 windings each, and are each symmetrically spaced from the LIF region with distances of 7.5(1)~cm and 11.3(1)~cm to the molecules.  

\subsection{Theory: Effective Hamiltonian}\label{sec:theory}

The ground and excited states are modeled with an effective Hamiltonian approach~\cite{Brown2003}. The \A{} state is well described by a Hund's case (a) Hamiltonian, using parameters from a previous optical study on a supersonic YbOH beam~\cite{Steimle2019}. Complete details of the effective Hamiltonian are provided in the supplementary materials. In the excited state, strong spin-orbit interactions mean $N$ is not a well-defined quantum number. Conversely, the molecule-frame projection quantum numbers $\Lambda$, $\Sigma$, and $\Omega$ are well-defined in Hund's case (a). Cross terms of spin-orbit and rotational perturbations give rise to the $\Lambda$-doubling interaction, which mixes the projection quantum numbers. The resulting Hund's case (a) $\tilde{A}$ eigenstates are symmetric and anti-symmetric superpositions of projections with well defined parity $\mathcal{P}$:

\begin{equation}
    |\Lambda; S,\Sigma;J,\Omega, M, \mathcal{P}=\pm\ket =\frac{1}{\sqrt{2}} (|\Lambda;S,\Sigma;J,\Omega,M\ket \pm (-1)^{p_a} |-\Lambda;S,-\Sigma;J,-\Omega,M\ket).
\end{equation}
The phase factor $p_a = J-S-\ell$ is connected to the convention for the action of the parity operator, $\mathcal{P}|\Lambda; S,\Sigma;J,\Omega,M\ket = (-1)^{p_a} |-\Lambda; S,-\Sigma;J,-\Omega,M\ket$. This phase convention is followed by Ref.~\cite{Hirota1985,Brown2000} (Details in the supplementary materials). 

We model the ground \Xbend{} state using a Hund's case (b) effective Hamiltonian describing a ${}^2 \Pi$ vibronic state. This approach has provided an accurate description of the vibrational bending modes in other metal hydroxide molecules, such as CaOH and SrOH in optical~\cite{Li1995} and millimeter wave~\cite{Fletcher1995} studies. The lack of first-order spin-orbit interaction means the electron spin $S$ is largely independent of the internuclear axis, and therefore both $\Sigma$ and $P$ are undefined. Hund's case (b) is the natural basis, with $N$ and its projection $\ell$ as good quantum numbers. The spin-rotation interaction then couples $N$ with $S$ to form well-defined $J$. Higher-order perturbations give rise to the $\ell$-doubling interaction, and the $\tilde{X}$ eigenstates of good parity are written as:

\begin{equation}
    |\ell; N,S,J,M, \mathcal{P}=\pm\ket =\frac{1}{\sqrt{2}} (|\ell; N,S,J,M\ket \pm (-1)^{p_b} |-\ell; N,S,J,M\ket).
\end{equation}
The phase factor in Hund's case (b) is defined as $p_b=(-1)^{N-\ell}$. The additional factor of $\ell=1$ means the action of the parity operator on a singly excited bending mode is similar to that of a $\Sigma^-$ electronic state. While this phase convention has physical basis (see supplementary materials) and has been used in literature~\cite{Hirota1985,Beaton1997,Brown2000,Allen2000CCN,Brown1975K}, the choice is not universal. The parity phase and the sign of the $\ell$-doubling Hamiltonian together determine if the lowest energy state is positive or negative parity. 

We use an effective Hamiltonian for the $\tilde{X}(010)$ state given by

\begin{equation}\label{eqn:Hbend}
    H_{\tilde{X}(010)} = B (\vec{N}^2-\ell^2) + \gamma (\vec{N}\cdot \vec{S} - N_z S_z) + \gamma_{G} N_z S_z 
    + \frac{p_G}{2} \left(N_+S_+ e^{-i 2 \phi} + N_-S_- e^{i2 \phi}\right) - \frac{q_G}{2} \left(N_+^2 e^{-i 2 \phi} + N_-^2 e^{i2 \phi}\right). 
\end{equation}

This form was first derived in Ref.~\cite{Merer1971} and is presented in detail in Refs.~\cite{Beaton1997,Allen2000CCN,Brown2003Rot}. Here, all subscripts on angular momenta ($z, \pm$) denote molecule-frame quantities. The azimuthal angle of the bending nuclear framework is given by $\phi$. The first term gives the rotational energy of a symmetric top. The next two terms describe the spin-rotation interaction coupling $N$ and $S$ to form $J$. The last two terms describe $\ell$-type parity doubling caused by terms off-diagonal in the vibrational angular momentum $G$, and cause splittings of opposite parity states. 

For the spin-rotation interaction we have modified the usual expression, $\gamma N \cdot S$, by subtracting $\gamma N_z S_z$ to account for the bending motion. This modification is crucial for accurate description of low-$N$ spectra (see supplementary materials). Other perturbations can reintroduce this axial spin-rotation term into the Hamiltonian, labeled in the literature with the coefficient $\gamma^\prime$~\cite{Merer1971} or $\gamma_G$~\cite{Beaton1997,Brown2003Rot}. The first order contribution to $\gamma_G$ arises from magnetic dipole interactions~\cite{Chang1970} and is negligible for the Yb-centered electron in YbOH. At higher order, a combination of vibronic coupling and spin-orbit interactions can contribute to $\gamma_G$ by mixing states with $\Pi$ electronic character, as observed in NCO~\cite{Gillett2006}, CCH~\cite{Carrick1983}, and FeCO~\cite{Tanaka2015}. 

In Eq. \ref{eqn:Hbend}, the $q_G$ parity-doubling term is standard for a bending molecule in a ${}^2 \Sigma$ electronic state. This term arises from Coriolis effects at second order, similar to the $q$ term in $\Lambda$-doubling. The $p_G$ term, also in analogy with $\Lambda$-doubling, is equivalent to a parity-dependent spin-rotation interaction. Owing to the weak coupling of the spin to the internuclear axis in $\Sigma$ electronic states, this term is small and has only been observed in submillimeter spectroscopy of metal hydroxides~\cite{Fletcher1995,Apponi1999MgOH}, ZnCN~\cite{Brewster2002ZnCN}, and CrCN~\cite{Flory2007CrCN}. As with $\gamma_G$, this term receives higher-order contributions from vibronic mixing with electronic $\Pi$ states. In spherical harmonic notation~\cite{Brown2003}, the $\ell$-type doubling terms may be written in the molecule frame as $\sum_{q=\pm1} e^{-2iq \phi} \left(p_G T^2_{2q}(N,S) - q_G T^2_{2q}(N,N)\right)$. 

We are using a sign convention for the $\ell$-type doubling Hamiltonian outlined by Brown~\cite{Brown1975K,Brown2003Rot}, where the $\ell$-type doubling Hamiltonian mirrors that used for $\Lambda$-doubling. However matrix elements of $\ell$ involve different phases than $\Lambda$. As a result of the $(-1)^\ell$ factor in our parity phase, we have the matrix elements $\bra \ell = \pm 1|e^{\pm 2 i \phi} | \ell^\prime=\mp1\ket = 1$, differing from the azimuthal matrix elements for $\Lambda$-doubling.  Matrix elements and complete details of the effective Hamiltonian and conventions used are provided in the supplementary materials.

We construct the predicted spectrum by first separately diagonalizing the effective Hamiltonians for the ground and excited states. The Hamiltonian basis is truncated at $N^{\prime\prime}=6$ for the \Xbend{} state and $J^\prime = 15/2$ for the \A{} state. Following Ref.~\cite{Steimle2019}, we include the $P=3/2$ manifold when diagonalizing \A{}. After obtaining eigenvectors and eigenvalues, we convert all eigenvectors to Hund's case (a) and compute matrix elements of the transition dipole moment (TDM) operator. Details of the TDM operator are given in section \ref{sec:intensity} and in the supplementary materials. For transitions with non-zero TDM, we compute the line position by taking the difference of excited and ground eigenvalues. 

\section{Results}\label{sec:results}

\subsection{Field-Free Spectrum}\label{sec:lines}

\begin{figure}[t]
    \resizebox{0.75\textwidth}{!}{\includegraphics{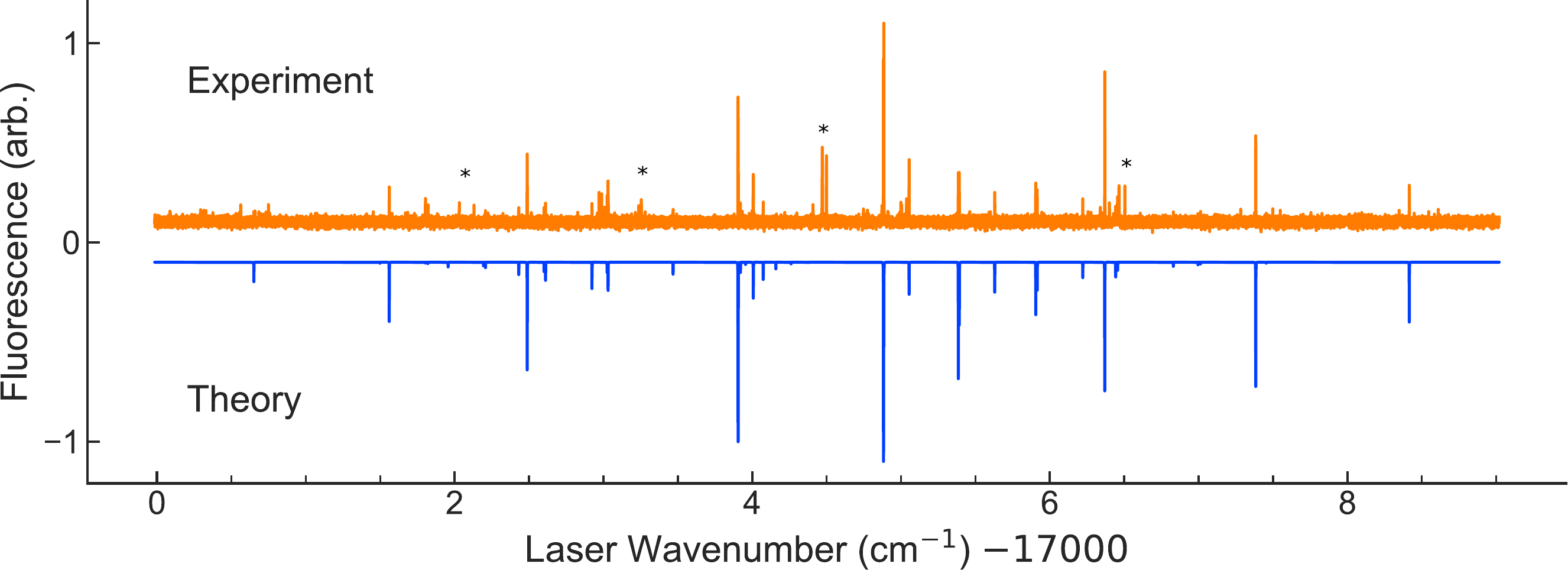}}
    \caption{\label{fig:spectrum} Field-free spectrum over a $\sim$9~cm$^{-1}$ range. Orange upper part is experimental observation and blue lower part is theory prediction. Prediction is using effective model detailed in section \ref{sec:intensity} with coefficients $(c_\mu=0.28,c_\kappa=-0.49,c_B=0.83)$ and a temperature of $T=2$ K. Lines marked with * are unassigned and could arise from other isotopologues or bands.}
\end{figure}

The observed spectrum (Fig \ref{fig:spectrum}) exhibits large splittings that match the excited state $\Lambda$-doubling and rotational separation. We perform combination-difference tests~\cite{Brown2003} with these splittings to obtain initial quantum number assignments of transitions. With these assignments, we compute initial guesses for the $B$, $\gamma$, and $q_G$ Hamiltonian parameters for the \Xbend{} state. Using these values and fixing the excited state parameters, we construct a predicted spectrum and perform further line assignments (line notation is described in \ref{sec:notation}). With this analysis, we determined the need for additional parameters $p_G$ and $\gamma_G$ to accurately describe the full spectrum.

Without the $p_G$ term, various $R$ and $P$ branch features deviate from the prediction by a magnitude $>$20~MHz, much larger than our frequency error. Specifically, in the region scanned in Fig.~\ref{fig:spectrum}, without $p_G$, lines with significant residuals are: ${}^R R_{11}^+(2), {}^R R_{11}^-(3)$, ${}^{O} {P}_{1{2}}^{{+}}({4})$, ${}^{P} {Q}_{1{2}}^{{+}}({5})$, and ${}^{P} {P}_{1{1}}^{{+}}({5})$. The magnitude and parity behavior of these residuals cannot be explained by centrifugal distortion, but can be explained by a parity-dependent spin-rotation interaction, namely $p_G$. By introducing $p_G$ into the prediction, all of these residuals are reduced to values commensurate with the experimental error. Furthermore, using the fit value of $p_G$, we predicted and found the ${}^{R} {R}_{1{1}}^{{+}}({4})$ and ${}^{R} {R}_{1{1}}^{{-}}({5})$ lines (not visible in Fig.~\ref{fig:spectrum}). These additional lines are added to the final fit and confirm the need for a $p_G$ term to accurately model the full spectrum. 

Unlike $p_G$, the $\gamma_G$ term does not scale with $N^{\prime\prime}$. However, we find this term necessary to describe the $N^{\prime\prime}=1$ structure, which was crucial for accurate Stark and Zeeman analysis in section \ref{sec:field}. In particular, we recorded multiple field-free calibration scans of the $^Q Q^+_{11}(1)$ and $^Q R^+_{12}(1)$ lines. Since these lines share the same excited state, their separation is insensitive to error in the \A{} state parameters. We use the separation of these lines to determine the $N^{\prime\prime}=1^+$ spin-rotation splitting to be 61.8(20)~MHz, and we add this value as an additional data point for our analysis. By including the $\gamma_G$ term in the spectral prediction, were we  obtain an accurate prediction of the $N^{\prime\prime}=1^+$ splitting commensurate with our measurement error. 

In total, we assigned 38 of the observed lines to 39 transitions originating from the $N^{\prime \prime}=1$ through $N^{\prime \prime} = 5$ levels of the \Xbend{} state. Note the ${}^Q R_{12}^{-} (1)$ and ${}^P Q_{12}^{-}(5)$ lines are overlapped. To obtain optimal effective Hamiltonian parameters, we vary the \Xbend{} state parameters and hold fixed the \A{} state parameters to the values given in Ref.~\cite{Steimle2019}. We construct predicted spectra and  perform nonlinear least-squares minimization of the residuals between the observed and predicted positions of all 39 assigned lines and the $N^{\prime\prime}=1^+$ spin-rotation splitting. A full list of line assignments is provided in the supplementary materials.

The best fit parameters are presented in Table~\ref{tab:params}. The fit residuals have a standard deviation of 6.1~MHz, consistent to order unity with the error reported in the previous optical study of the \A{} state~\cite{Steimle2019}.  The rotational and spin rotational $\tilde{X}(010)$ parameters are in good agreement with those for $\tilde{X}(000)$ and $\tilde{X}(100)$, also collected in Table~\ref{tab:params}. The location of the origin $T_0$ is in excellent agreement with previous dispersed fluorescence studies~\cite{Mengesha2020YbOH,Zhang2021VBR}. The rotational constant $B$ decreases in \Xbend{} as a result of vibrational corrections. The increasingly negative spin-rotation parameter $\gamma$ between the three vibrational states is a result of second order spin-orbit perturbations from low-lying electronic states with 4\textit{f}$^{13}$6\textit{s}${}^2$ electronic configuration for the Yb centered electron, known as ``4f hole'' states~\cite{Nakhate2019,Zhang20224f}.

Vibronic mixing with electronic ${}^2 \Pi$ states can also explain the observed $\gamma_G$ and $p_G$ parameters, which are not typical for the bending mode of an isolated electronic ${}^2 \Sigma$ state. Vibronic mixing exchanges $\ell$ and $\Lambda$ while preserving $K$. As a result, the \Xbend{} state can acquire some $\Lambda>0$ electronic character, inheriting spin-orbit and $\Lambda$-doubling interactions from neighboring ${}^2 \Pi$ states. Specifically, in the effective Hamiltonian, these interactions can arise at third-order via a combination of linear vibronic coupling and spin-orbit effects. This term was first described by Brown in the context of spin-orbit corrections to electronic ${}^2 \Pi$ states as a result of mixing with other ${}^2 \Sigma$ or ${}^2 \Delta$ states~\cite{brown1977effective}. Neighboring states that can contribute to $\gamma_G$ and $p_G$ include both the $\tilde{A}$ manifold and the 4\textit{f} hole states.  The exact nature of the 4\textit{f} hole states and their vibronic mixing in YbOH is currently unknown and merits further study. However, their proximity to the ground state and their large spin-orbit interactions could explain the significant magnitude of $p_G$ and $\gamma_G$ in YbOH compared to other metal hydroxides~\cite{Fletcher1995}. 

The $\ell$-type doubling parameter $q_G$ is a similar magnitude to that of other metal-hydroxide \Xbend{} states~\cite{Li1995,Fletcher1995}, and is in agreement with a recent theoretical calculation~\cite{Zakharova2022}. The parameter $q_G$ can be interpreted in terms of the Coriolis coupling constants of a triatomic molecule~\cite{Nielsen1951,Li1995}:

\begin{equation}
    q_G=-(v_2+1)\frac{B^2}{\omega_2}\left(1+\sum_{n=1,3} \zeta_{2n}^2 \frac{4\omega_2^2}{\omega_n^2-\omega_2^2}\right).
\end{equation}
Here, $v_2$ is the number of quanta in the bending vibration $\omega_2$, and $\zeta_{2n}$ is the Coriolis coupling constant between the bending mode and the $v_n$ stretch modes. To estimate $\zeta_{21}$, we can estimate the value of $\omega_3$ (O-H stretch) using the CaOH value of 3778~cm$^{-1}$~\cite{Pereira1996}, and we set $v_2=1$, $\omega_2 \approx T_0$, and $\omega_1\approx529.3$~cm$^{-1}$~\cite{Steimle2019}. Furthermore, we can use the relationship $\zeta^2_{21}+\zeta^2_{23}=1$~\cite{Nielsen1951} to eliminate $\zeta^2_{23}$. Using our values of $B$ and $q_G$, we then obtain a value of $\zeta_{21}\approx 0.137$, slightly smaller than in CaOH (0.1969)~\cite{Li1995} and SrOH (0.179)~\cite{Presunka1993}. This is likely due to the break down of the harmonic approximation $\omega_2\approx T_0$ and the approximation of $B_e\approx B$. Further work is needed for a complete vibrational characterization. 

\begin{table*}[t]
\centering
\begin{threeparttable}
\caption{\label{tab:params}
Spectroscopic parameters for the low-lying vibrational states of the $\tilde{X}{}^2\Sigma^+$ manifold. The \Xbend{} parameters are obtained from the current work. }
\begin{ruledtabular}
\begin{tabularx}{0.75\textwidth}
{l ddd}

Parameter & \multicolumn{1}{c}{$\tilde{X}(000)$~\protect\cite{Nakhate2019}} & \multicolumn{1}{c}{\Xbend{}} & \multicolumn{1}{c}{$\tilde{X}(100)$~\protect\cite{Steimle2019}}  \\ \hline 
$T_0$/cm$^{-1}$ & \multicolumn{1}{c}{0} & \multicolumn{1}{c}{319.90901(6)} & \multicolumn{1}{c}{529.3269(3)}  \\
$B$/MHz & 7348.4005(3) & 7328.64(15) & 7305.37(24)\\
$\gamma$/MHz & -81.15(6) & -88.7(9) & -110.6(21) \\
$\gamma_G$/MHz & \multicolumn{1}{c}{--} & \multicolumn{1}{c}{$\,\,$16(2)} & \multicolumn{1}{c}{--} \\ 
$q_G$/MHz & \multicolumn{1}{c}{--} & -12.0(2) & \multicolumn{1}{c}{--}\\
$p_G$/MHz & \multicolumn{1}{c}{--} & \multicolumn{1}{c}{$-$11(1)$\,\,$} & \multicolumn{1}{c}{--}\\

\end{tabularx}
\end{ruledtabular}
\end{threeparttable}
\end{table*}

\begin{figure}[ht]
    \resizebox{0.5\linewidth}{!}{\includegraphics{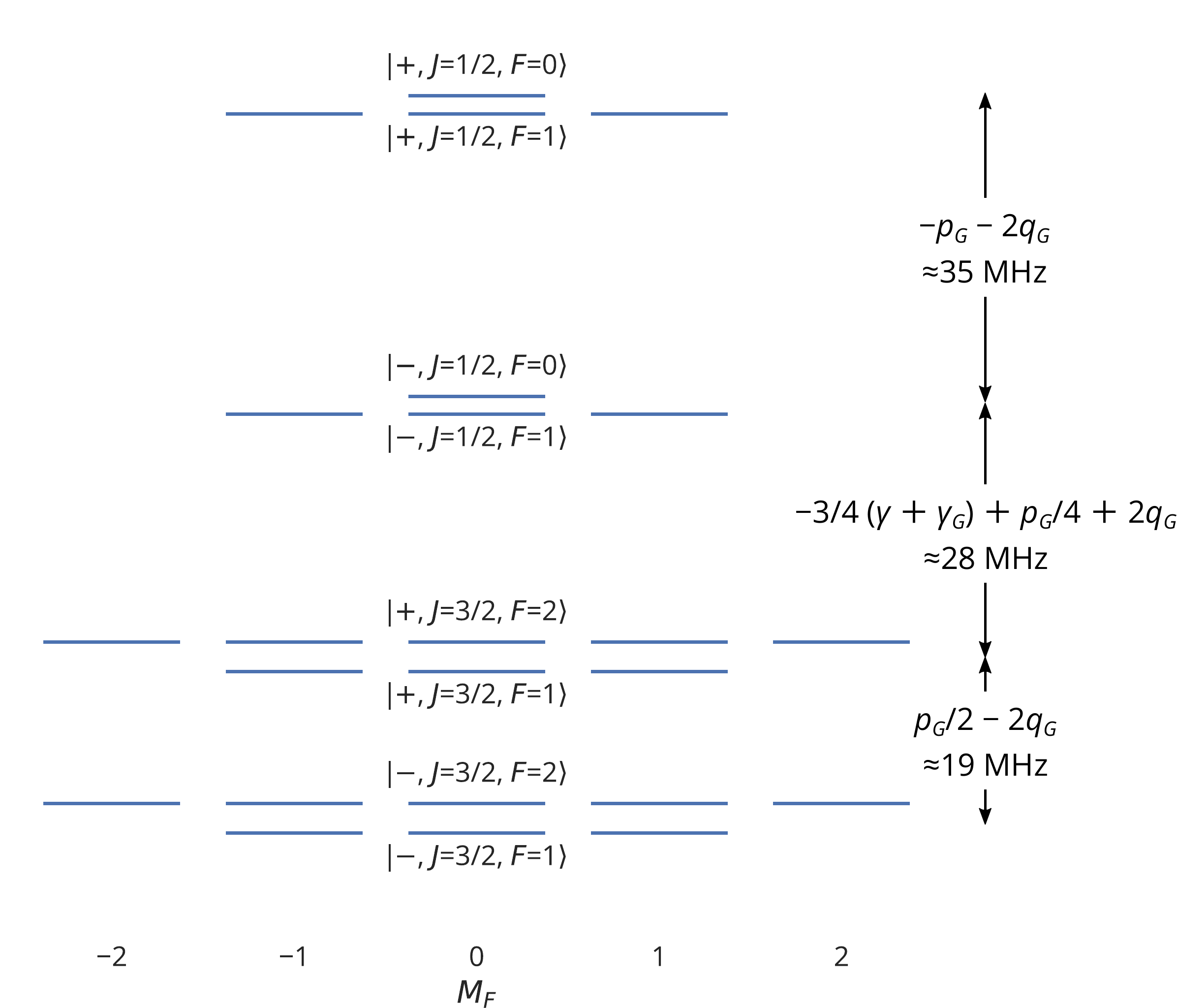}}
    \caption{\label{fig:structure} Field-free level structure of the $N=1$ manifold in the \Xbend{} state. States are arranged vertically by energy and horizontally by their $M_F$ angular momentum projection. States are labeled in the parity basis.  The hyperfine structure was not resolved in our work, and is instead approximated using parameters from a study of the \X{} state~\protect\cite{Nakhate2019}. }
\end{figure}

Using the parameters obtained from our analysis, we construct a field-free level diagram for the $N=1$ manifold of the \Xbend{} state, shown in Figure~\ref{fig:structure}. As stated previously, $N=1$ is the lowest rotational manifold in the \Xbend{} state, as we always have $|\vec{N}\cdot\hat{n}|=1$. Due to their small parity splittings, $N=1$ states are easily polarized, making them useful for precision measurements~\cite{Kozyryev2017PolyEDM}. The effect of the parity-dependent spin-rotation term, $p_G$, is apparent in the asymmetric parity-doubling of the $J=1/2$ and $J=3/2$ manifolds. Though we are not sensitive to hyperfine splittings, for completeness we have included the H hyperfine structure using the parameters obtained for the \X{} state in a previous study~\cite{Nakhate2019}. The hyperfine structure is not expected to change significantly in the bending mode. 

The recorded spectrum has lines present that could not be assigned with combination-differences using the $\tilde{A} (000)$ structure, and are not observed in the prediction using the best-fit parameters. The lines are marked with * in Fig.~\ref{fig:spectrum}. We conclude that some of these lines are indeed from ${}^{174}$YbOH by comparing their chemical enhancement~\cite{Jadbabaie2020} when using ${}^1S_0\rightarrow{}^3P_1$ transitions for different Yb isotopes. These lines could be unthermalized rotational states, or possibly another overlapping $\Delta \ell = \pm 1$ band, such as the $\tilde{X} {}^2 \Sigma^+ (02^{0,2}0) \rightarrow \tilde{A} {}^2 \Pi_{1/2} (010)$ bands.

\subsection{Stark and Zeeman Spectra}\label{sec:field}

After fitting the molecular structure with the field-free spectrum, we study the Stark and Zeeman spectra of the molecule in the presence of static (DC) electric and magnetic fields, using the experimental setup described in \ref{sec:methods}.  We obtain the spectra by scanning the 588~nm probe laser across two lines corresponding to the field-free $N^{\prime\prime} = 1^+ \rightarrow J^\prime = {\frac{3}{2}}^-$ transition, $^Q Q^+_{11}(1)$ and $^Q R^+_{12}(1)$. The applied DC fields point along $z$, while the laser polarization is along $x$. Spectra are taken with the E-field varied from $0 - 264$~V/cm and with the applied B-field varied from $0 - 70$~G. Calibration spectra are taken with $E_Z=0$~V/cm and $B_Z<0.5$~G, and the observed line positions are compared to the I$_2$-corrected field-free positions to calibrate for frequency offsets. 

The lines of interest are relatively well-isolated from other features, and the small $N^{\prime\prime}=1$ parity doubling allows us to enter the linear stark regime with modest laboratory fields $\gtrsim$100~V/cm. Since the parity splittings of the excited $\tilde{A} {}^2 \Pi_{1/2}$ state are $>$13 GHz, and its molecule frame dipole moment is $D_\mathrm{\tilde{A}} = 0.43(10)$ D~\cite{Steimle2019}, at the fields we consider the excited state Stark shifts are essentially negligible. Furthermore, given our frequency resolution and the natural linewidth, we are only sensitive to the isotropic interaction of $B_Z$ with the electron spin magnetic moment. Curl-type relationships~\cite{Allen2000CCN} estimate anisotropic spin interactions at $6\times10^{-3}\mu_B$, and the nuclear magnetic moment is also suppressed at a similar level, with both effects giving shifts below our resolution.

\begin{figure}[t]
    \centering
    \resizebox{0.75\linewidth}{!}{\includegraphics{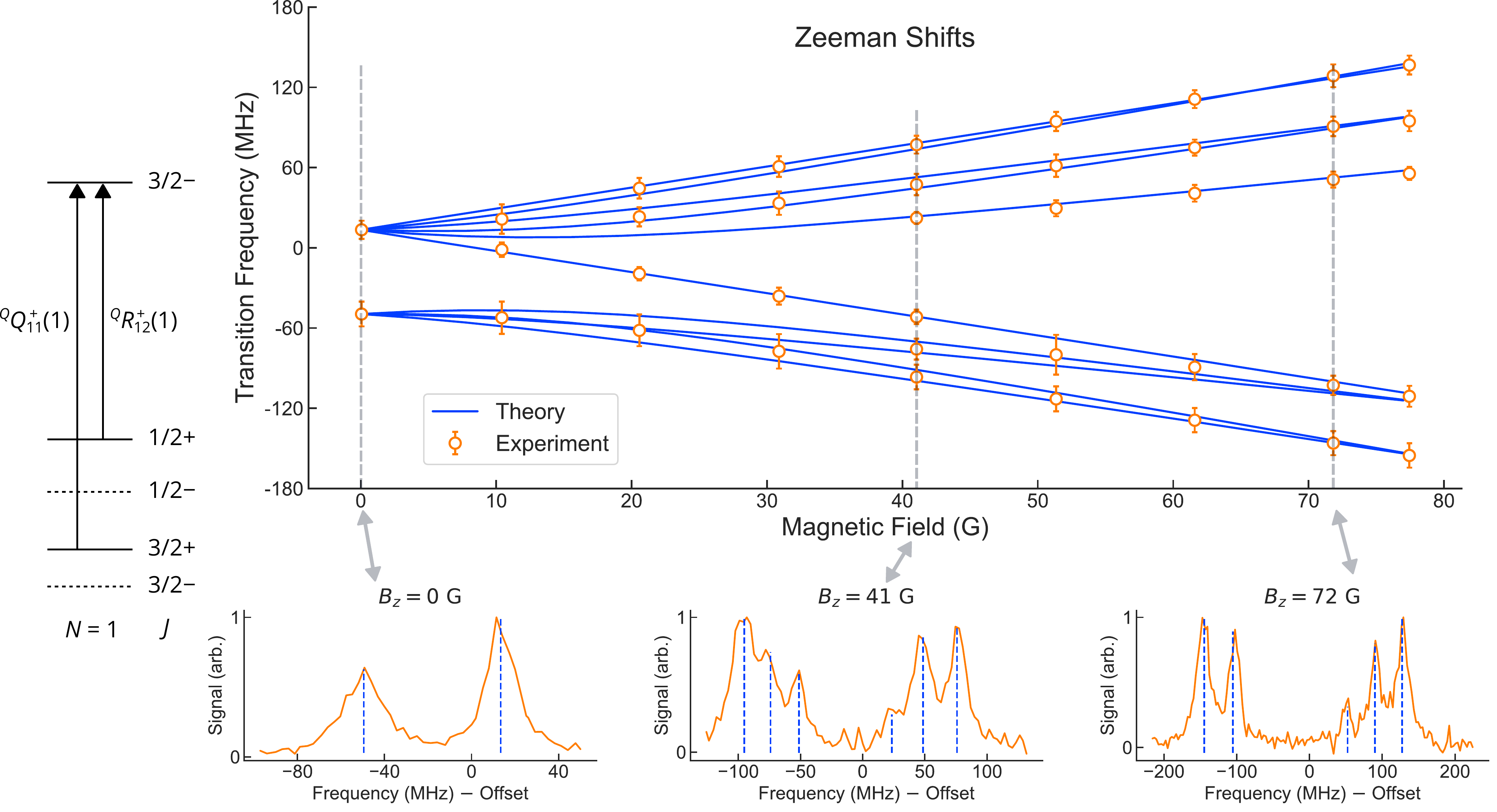}}
    \caption{\label{fig:zeeman} Zeeman spectroscopy of the \Xbend~state. The main plot shows the transition frequency shift (with subtracted offset) in a magnetic field, the blue lines are optimized model predictions, and the orange circles are experimental measurements. Error bars are 1-$\sigma$ measured peak widths, set by a combination of radiative broadening and unresolved hyperfine structure, limiting the ability to resolve closely-spaced lines. Lower subplots are slices of the spectra at various magnetic field values, with experimental data in orange and predicted line locations indicated with vertical dashed blue lines. On the left, we show the field-free level structure of the transitions studied.}
\end{figure}

\begin{figure}[t]
    \centering
    \resizebox{0.75\linewidth}{!}{\includegraphics{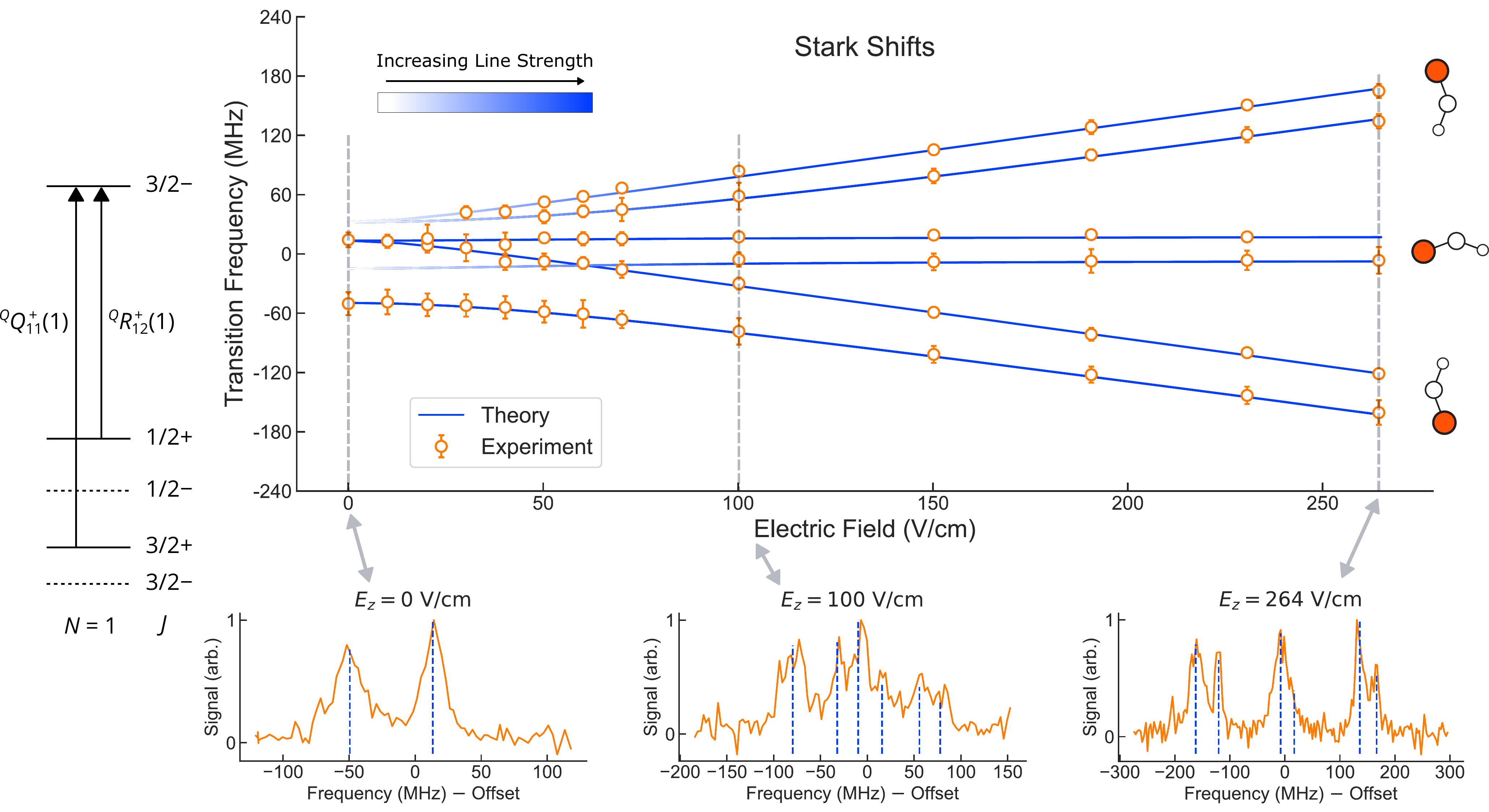}}
    \caption{\label{fig:stark} Stark spectroscopy of the \Xbend~state. The main plot shows the transition frequency shift (with subtracted offset) in an electric field, the blue lines are optimized model predictions, and the orange circles are experimental measurements. The blue color gradient represents parity forbidden transitions that gain strength at finite electric field. Error bars are 1-$\sigma$ peak widths, set by a combination of radiative broadening and unresolved hyperfine structure, limiting the ability to resolve closely-spaced lines. Lower subplots are slices of the spectra at various electric field values, with experimental data in orange and predicted line locations indicated with vertical dashed blue lines. On the left, we show the field-free level structure of the transitions studied.}
\end{figure}

To obtain energy levels and predicted lines, we fix the field-free parameters and diagonalize the combined Stark, Zeeman, and field-free Hamiltonian. We obtain optimal estimates for free Stark and Zeeman parameters by least-squares minimization of the residuals between observed and predicted line positions. 

Both ground and excited levels are magnetically sensitive. The Zeeman shifts of the $\tilde{A} {}^2 \Pi_{1/2} (000)$ and $\tilde{X} {}^2 \Sigma^+ (000)$ states were previously studied at similar magnetic field strengths in Ref.~\cite{Steimle2019}, and recently at high fields ($\sim$1~T) in Ref.~\cite{ZS2022}. Following these references, we use the following effective Zeeman Hamiltonians for the ground and excited states:
\begin{subequations}
\begin{align}
    & H^{Zee}_{X} = g_S \mu_B S_Z B_Z \\
    & H^{Zee}_{A} = g^\prime_S \mu_B S_Z B_Z + g_L L_Z B_Z + g^\prime_l \mu_B \left(e^{-2i\theta} S_+ B_+ + e^{2i\theta} S_- B_-\right)
\end{align}
\end{subequations}
Here, $Z$ refers to the lab-frame projection, $\pm$ refer to the molecule frame projections, and $\theta$ is the electronic azimuthal coordinate. For the excited state, we use the values from Ref.~\cite{ZS2022}, fixing $g^\prime_S=1.860$, $g_L=1.0$, and $g^\prime_l = -0.724$. For the ground state, we allow $g_S$ to vary in the fits to find an effective value that accurately describes the Zeeman shifts. While we do not include them here, at higher resolution or at higher field values, additional terms are expected to contribute in the effective Zeeman Hamiltonian, including terms associated with the bending angular momentum~\cite{Allen2000CCN}. 

The Zeeman fits prefer a value of $g_S = 2.07(2)$, deviating from the free electron \textit{g}-factor of 2.0023. The experimental Zeeman shifts and the prediction from the optimized model are shown in Fig.~\ref{fig:zeeman}. Corrections to $g_S$ can arise from mixing involving other states with different Zeeman tuning. For example, the Zeeman shifts of the \A{} state were fit to $g^\prime_S=1.860$ in a recent high-field study~\cite{ZS2022}, owing to perturbing 4\textit{f}$^{13}$6\textit{s}${}^2$ states. Since we observe perturbations from these 4\textit{f} states in the field-free structure of the \Xbend{} state, it is natural to also find their effects in the Zeeman shifts. Furthermore, the 4\textit{f} states are split into a higher energy spin-orbit anti-aligned manifold and a lower energy spin-orbit aligned manifold~\cite{Zhang20224f}. Due to energy proximity, while \A{} predominantly interacts with the 4\textit{f} hole anti-aligned manifold, \Xbend{} will be perturbed more strongly by the aligned manifold. The difference in electron orientation of the two spin-orbit 4\textit{f} manifolds can explain the difference between \Xbend{} and \A{} in the sign of the deviation of $g_S$ from its nominal value.

To describe the Stark shifts, for the both ground and excited states we use the Hamiltonian $H_E = -\vec{D}_\mathrm{mol}\cdot \vec{E}$. The molecule frame dipole moment $D_\mathrm{mol}$ is kept as a free parameter, and obtained from spectra where $E_Z$ is scanned with $B_Z<0.5$~G. The optimal fit value is $D_\mathrm{mol} = 2.16(1)$~D = $1.09$ $h$~MHz/(V/cm). This value is in good agreement with the measured $\tilde{X} (000)$ dipole moment of 1.9(2)~D. In Figure \ref{fig:stark}, we plot the theoretical prediction based on the optimal fit against the observed line positions. 

The Stark shifts confirm the assignment of the $\tilde{X}(010)$ state and demonstrate the orientation control afforded by parity doublets. In the bending mode, the projection of the molecular axis on the lab-frame $Z$-axis is given by $\hat{n}\cdot \hat{Z}= \frac{(\vec{N}\cdot \vec{Z})(\vec{N}\cdot \hat{n})}{N(N+1)} \propto M_N \ell$. Note we use $X, Y, Z$ to denote lab-frame axes and $x, y, z$ to denote the molecule-frame.  The molecule $z$ axis and dipole moment $D_\mathrm{mol}$ both point from O to Yb. For field-free states, $\langle M_N\ell\rangle = 0$, and the molecule is unpolarized. In the presence of an electric field fully mixing parity doublets, the Stark shifts are linear, and the eigenstates are diagonal in the the decoupled basis $|\ell;M_N, M_S\ket$. In this regime, the levels split into $2N+1$ dipole moment orientations pointing along $\frac{M_N \ell}{N(N+1)}$, and splittings within each orientation manifold are due to the spin-rotation interaction. 

\subsection{Anomalous intensities and perturbations}\label{sec:intensity}

Since the \A{} state has been previously fully characterized~\cite{Steimle2019}, the assignment of energy levels in \Xbend{} is fairly straightforward using the effective Hamiltonian approach.  However, because this transition is nominally forbidden, interpreting the line intensities is a challenge. Electric dipole (E1) transitions involving $\Delta \ell \neq 0$ are forbidden in the Condon approximation, which separates electronic and vibrational degrees of freedom~\cite{Herzberg1967,Demtroder2005}. These nominally forbidden vibronic transitions have been observed spectroscopically in many species of linear triatomic molecules, including NCO~\cite{Bolman1973NCO}, NCS~\cite{Dixon1968NCS}, MgNC~\cite{Fukushima2007MgNC}, CaOH~\cite{Jarman1992,Coxon1994,Li1995}, SrOH~\cite{Presunka1993,Presunka1994,Lasner2022}, and YbOH~\cite{Mengesha2020YbOH}, though modeling of the intensities is less common.

These transitions borrow intensity from E1-allowed bands through a combination of vibronic and spin-orbit perturbations~\cite{Baum2021cycling,Zhang2021VBR}. Branching ratios involving forbidden vibronic transitions in YbOH were measured in a previous study~\cite{Zhang2021VBR} examining dispersed fluorescence from the $\tilde{A} (000)$ state, with resolution at the $10^{-5}$ level. The experimentally observed vibrational branching was in good agreement with a theoretical study published in the same work~\cite{Zhang2021VBR}. While these transitions are of interest as leakage channels for photon cycling, they can also be a resource for spectroscopy, as we show in the current work. 

The observed spectrum exhibits anomalous rotational line intensities, with certain transitions completely missing at our level of sensitivity. For example, despite their expected thermal occupation ($N^{\prime\prime}\leq3$), the ${}^P Q^+_{12}(1)$, ${}^P P^+_{11}(2)$, ${}^Q Q^+_{11}(2)$, ${}^P P^-_{11}(3)$, ${}^Q P^-_{11}(3)$, and ${}^Q R^-_{12}(3)$ lines are missing (see Supplemental Material for a full list of lines). Anomalous line intensities for forbidden transitions have been previously observed in other molecules with vibronic mixing~\cite{Fukushima2007MgNC,Li1995,Presunka1993,Presunka1994,Coxon1994}. By considering the intensity-borrowing that gives transition strength to these forbidden transitions, we develop a model that qualitatively explains the observed line strengths.

In an E1 transition, the transition strength is proportional to the square of the transition dipole moment between the ground and excited state, $|\bra \tilde{A}|T^1_p(d)|\tilde{X}\ket|^2$. We are using spherical tensor notation, where $p$ denotes the component of the spherical tensor in the lab-frame and $q$ in the molecule-frame. Using a Wigner $\mathcal{D}$ matrix, we can write the lab frame dipole moment in terms of its molecule frame projections: $T^1_p(d) = \sum_q \mathcal{D}^{(1)}_{p,q}(\omega)^* T^1_q(d)$. In the E1 approximation, $\Delta \Sigma = 0$, and the molecule-frame projection $q$ of the transition dipole moment determines the selection rule for $\Lambda$. The perpendicular $q=\pm 1$ components drive $\Delta \Lambda = \pm 1$ transitions, for example the allowed $\tilde{A} - \tilde{X}$ band, while parallel $q=0$ component drives $\Delta \Lambda = 0$, for example the allowed $\tilde{B} - \tilde{X}$ band.

In the limit of very large vibronic interaction, $\Lambda$ and $\ell$ are fully mixed, and one might consider the $\tilde{X}(010)\rightarrow \tilde{A}(000)$ transition as a vibronic ${}^2\Pi - {}^2\Pi$ parallel band, with $\Delta K=0$. In reality, the vibronic mixing is perturbative in the ground and excited states, and $\Lambda$ and $\ell$ are well-defined. As a result, the observed line intensities are completely inconsistent with a solely parallel transition model.

Instead, we model the $\tilde{X}(010) \rightarrow \tilde{A}(000)$ transition as a mixture of perpendicular and parallel bands. We consider the effects of vibronic perturbations with the selection rule $\Delta \ell = \pm 1$, which can result in intensity borrowing. At first order, we have the dipolar Renner-Teller (RT) Hamiltonian, also referred to as Herzberg-Teller coupling~\cite{Brown2000,Hirota1985,Bolman1973NCO},
\begin{equation}
    H_{RT} = \frac{V_{11}}{2}\left( L_+ q_- e^{i(\theta-\phi)} + L_- q_+ e^{-i(\theta-\phi)}\right).
\end{equation}
This interaction is a form of linear vibronic coupling~\cite{Koppel1981}. Here, $V_{11}$ parameterizes the interaction strength, $\theta$ is the electronic azimuthal coordinate, $\phi$ is the bending azimuthal coordinate as before, $L_\pm$ is a raising/lowering operator with $\Delta \Lambda =\pm1$, and $q_\pm$ is a dimensionless raising/lowering operator with $\Delta \ell = \pm1$. Physically, this interaction can be interpreted as the electrostatic interaction between the displaced bending dipole moment and the electron cloud. The interaction preserves the composite projection number $K = \Lambda + \ell$.

At second order, the dipolar RT Hamiltonian can combine with the perpendicular spin-orbit Hamiltonian, 
\begin{equation}
    H_{SO} = \frac{A_\perp}{2} \left(L_+ S_- + L_- S_+\right),
\end{equation}
Where $L_\pm$ is defined as before, $A_\perp$ is the off-diagonal spin-orbit coupling, and $S_\pm$ is the raising/lowering operator with $\Delta \Sigma = \pm 1$. The combination of $H^{(1)}_{RT} \times  H^\perp_{SO}$ is an effective interaction with terms $q_\pm S_\mp$. This interaction has $\Delta K = - \Delta \Sigma = \pm 1$, but preserves the total angular momentum projection number $P = \Lambda + \Sigma + \ell$.  

\begin{figure}[t]
    \centering
    \includegraphics[width=0.5\textwidth]{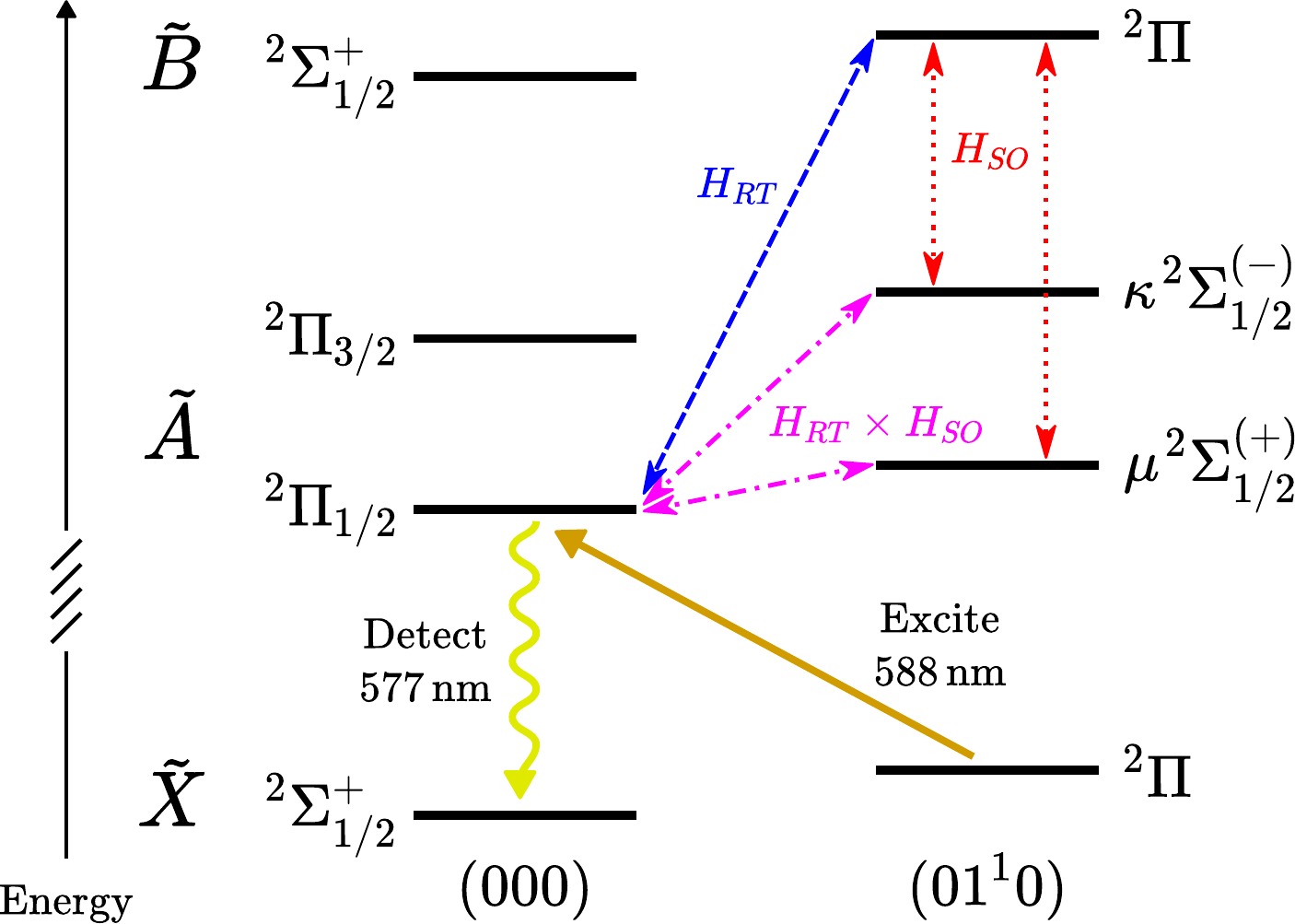}
    \caption{\label{fig:pert} Level schematic for relevant states and perturbations in YbOH. Levels are labeled by their vibronic term symbol. We detect the $\tilde{X}(010)$ bending state (which is a vibronic ${}^2\Pi$ state) by laser excitation (orange line) up to the $\tilde{A} {}^2 \Pi_{1/2} (000)$ state and observe the fluorescence from decays to the ground $\tilde{X}(000)$ state (yellow wavy line). This excitation is a forbidden E1 transition, however, it acquires intensity by mixing of the excited $\tilde{A} {}^2 \Pi_{1/2} (000)$ state with other $|\ell|=1$ states. Mixing with $\tilde{B}(010)$ occurs via first-order (blue) Renner-Teller (RT) interactions, and mixing with the $\mu, \kappa (010)$ states occurs via second-order (purple) cross terms between RT and spin-orbit (SO) (red) interactions. Not shown for simplicity are similar SO interactions between $\tilde{A} {}^2 \Pi_{1/2} (000)$ and $\tilde{B} (000)$ and similar RT interactions between $\mu, \kappa (010)$ and $\tilde{B} (000)$, which also contribute to state mixing.}
\end{figure}

Denote the unperturbed excited state as $|\tilde{A} {}^2 \Pi_{1/2} (000)\ket_0$ and the true, perturbed eigenstate as $|\tilde{A} {}^2 \Pi_{1/2} (000)\ket$. We can then expand the perturbed eigenstate in terms of dominant $\ell = 1$ vibronic contributions~\cite{Zhang2021VBR,Baum2021cycling}:

\begin{equation}
     |\tilde{A} {}^2 \Pi_{1/2} (000)\ket \propto |\tilde{A} {}^2 \Pi_{1/2} (000)\ket_0  + c_{\mu} | \mu {}^2 \Sigma^{(+)}_{1/2}(010)\ket_0 + c_{\kappa} | \kappa {}^2 \Sigma^{(-)}_{1/2}(010)\ket_0 + c_{B} | \tilde{B} {}^2 \Pi (010)\ket_0.
\end{equation}
The perturbative coefficients $c_\mu,c_\kappa,c_B$ represent the relative admixture of the intensity-borrowing states. The relevant states and perturbations are shown schematically in Fig.~\ref{fig:pert}. The $\mu {}^2 \Sigma^{(+)}_{1/2}$ state is the $P=1/2$ component of the $\Omega=1/2$, $v_2=1, \tilde{A}$ manifold, and the $ \kappa {}^2 \Sigma^{(-)}_{1/2}$ state is the $P=1/2$ component in the $\Omega=3/2$, $v_2 = 1, \tilde{A}$ manifold. These two states are connected to $\tilde{A} {}^2 \Pi_{1/2} (000)$ by the second-order perturbation $H_{RT} \times  H_{SO}$. The $\tilde{B} {}^2 \Pi$ vibronic state is the $v_2=1$ component of the $\tilde{B} {}^2 \Sigma^+_{1/2}$ electronic state, and is connected to $\tilde{A} {}^2 \Pi_{1/2} (000)$ state via the first-order perturbation $H_{RT}$. 

Each of these perturbing states contribute to different molecule-frame components of the transition dipole moment (TDM). For example, the transition $\tilde{X} {}^2\Pi \rightarrow \tilde{B} {}^2 \Pi$ is generated by the $q=0$, $z$ component of the TDM, with $\Delta K = \Delta P =0$. The other transitions to $\mu$ and $\kappa$ have $\Pi\rightarrow\Sigma$ vibronic character, and couple via the $q=\pm 1$, $x,y$ TDM components. The perturbing $\mu$ and $\kappa$ states have opposite spin orientation compared to the original $\tilde{A}{}^2\Pi_{1/2}$ state. This means the intensity-borrowing states have mixed spin projection $\Sigma$, and the $\Delta \Sigma=0$ selection rule is not well-defined. 

The transition was modeled by first diagonalizing the $\tilde{A} {}^2 \Pi_{1/2}(000)$ and $\tilde{X} {}^2 \Sigma (010)$ states separately to obtain the level positions of both states. To evaluate the TDM, the excited state vector is then replaced by a linear combination of the intensity-borrowing state vectors with coefficients $c_\mu, c_\kappa, c_B$. The change of basis from $\tilde{A}$ to $\mu, \kappa$, and $\tilde{B}$ uses appropriate selection rules for vibronic mixing and preserves parity (see supplementary materials for details). The total TDM is the sum over the individual TDMs evaluated between \Xbend{} and the intensity-borrowing states. To obtain the transition intensity, the TDM is squared after the sum, allowing TDMs from different states to interfere with each other. This interference is the source of the anomalous line intensities.

The mixing coefficients, $c_\mu, c_\kappa, c_B$ could not be modeled with a deperturbation Hamiltonian, since neither the $\mu$, $\kappa$, or $\tilde{B}$ state have been extensively studied or modeled, and both states are expected to be affected by perturbations from nearby states with 4\textit{f}$^{13}$6\textit{s}${}^2$ Yb character~\cite{Zhang20224f}. Instead, the mixing coefficients are kept as free parameters and their ratios were fit to the experimentally observed, relative field-free intensities. For the intensity fits, the rotational temperature is fixed at $T=2$ K (the molecule beam is cooled by expansion out of the cell aperture), and since only relative intensities were fit, the $c_B$ parameter is held fixed. The normalized best fit mixing coefficients are found to be $(c_\mu, c_\kappa, c_B) = (0.28, -0.49, 0.83)$. This implies $\sim$69$\%$ of the $\ell=1$ character in \APi{} arises from mixing with $\tilde{B}(010)$, $\sim$24$\%$ from $\kappa (010)$, and $\sim$7$\%$ from $\mu (010)$. This is in good agreement with recent theory work on intensity borrowing in YbOH, which attributed 70$\%$ of the intensity borrowing to mixing with $\tilde{B}(010)$~\cite{Zhang2021VBR}. However, it is important to note that due to interference effects, relative amplitudes of the coefficients, not their squares, are important for determining rotational line intensities.

We find that using these parameters to model the transition provides good qualitative understanding of the observed spectrum, as evidenced by the theory and experiment comparison in Figure \ref{fig:spectrum}. Further studies of the excited state perturbations would be required to improve the fit; however, as the exact intensities are not critical for future experiments with this molecule, this model is sufficient to provide physical understanding of the intensities and behavior of this transition.

\vspace{-2mm}

\section{Conclusion}\label{sec:conclusion}

In this work, we performed high-resolution optical spectroscopy on the rovibrationally forbidden \XSigbend{}${}\rightarrow{}$\APi{} transition of ${}^{174}$YbOH. In total, we observed 39 transitions out of low rotational states with $N^{\prime\prime}\leq 5$. The \Xbend{} structure is well-described by a Hund's case (b) ${}^2\Pi$ effective Hamiltonian, and the $\ell$-type parity doubling is described by two constants, $q_\ell = -12.0(2)$~MHz and $p_\ell =-11(1)$~MHz. We modeled the anomalous line intensities of the forbidden band with mixing coefficients representing vibronic perturbations in the excited state. The anomalous intensities arise from quantum interference between TDMs from the perturbing $\tilde{B}(010)$, $\mu(010)$, and $\kappa(010)$ states. From the Zeeman spectra, we found the magnetic tuning of \Xbend{} is consistent with an effective isotropic electron spin \textit{g}-factor, $g_S=2.07(2)$. From the Stark spectra, we extracted the molecule-frame dipole moment of 2.16(1) D. These values are in good agreement with the parameters of the \X{} state. 

In our study, the hyperfine structure and higher-order Zeeman \textit{g}-factors were unresolved. Our work provides a basis for future studies with narrow-linewidth methods, such as RF, microwave, and two-photon spectroscopy, to precisely determine these properties. 

This work is an essential step towards measurements of CP-violating physics in YbOH~\cite{Kozyryev2017PolyEDM}, as well as other metal hydroxide molecules proposed for CP violation and parity violation searches that utilize the parity doublets in the bending mode.  We showed the \Xbend{} state $\ell$-doubling offers spectroscopically resolvable states of molecule polarization pointing along, against, and perpendicular to the applied electric field, over a wide range of field values. This orientation control over the dipole moment offers robust systematic error rejection without compromising laser cooling. The combination of these features make linear polyatomics a promising platform for new physics searches. With our measured data, we can compute the EDM sensitivity, which is proportional to the electron spin projection on the internuclear axis, $\Sigma$. We find a local maximum value of $\langle\Sigma\rangle=0.40$ in the $N=1, J=\frac{1}{2}^+$ state at $E=101$~V/cm, similar to what was predicted in prior theoretical work~\cite{Petrov2022,Augenbraun2021Thesis}. Furthermore, understanding the structure of ${}^{174}$YbOH is a step toward characterizing the more complicated structure of the odd isotopologues ${}^{171}$YbOH and ${}^{173}$YbOH, which have sensitivity to parity violation~\cite{Norrgard2019} and hadronic CP violation~\cite{Flambaum2014}, respectively. 

Lastly, our determination of the \Xbend{} location and structure is crucial for understanding the complicated excited state structure in YbOH. For example, with our knowledge of the bending frequency, we can tentatively assign the unknown [17.33] band in Ref.~\cite{Mengesha2020YbOH} to the \XSigbend{}${}\rightarrow\tilde{A} {}^2 \Pi_{1/2} (010)$ band. This would put the excited $\tilde{A} {}^2 \Pi_{1/2} (010)$ manifold at approximately 17652~cm$^{-1}$. This state is an excellent candidate for optically pumping population from \X{} into \Xbend{}, an important step for signal-to-noise-ratio improvements in precision measurements using the bending mode. Furthermore, the location of \Xbend{} is necessary for the determination of repumping pathways for laser cooling, slowing, and trapping of YbOH, toward next-generation CP violation searches. 

\vspace{-3mm}

\begin{acknowledgments}

\vspace{-2mm}

We acknowledge many helpful discussions with the PolyEDM collaboration and the Doyle group at Harvard. We thank Tim Steimle, Phelan Yu, and Ashay Patel for helpful discussions.  This work was supported by a NIST Precision Measurement Grant (60NANB18D253), an NSF CAREER Award (PHY-1847550), the Heising-Simons Foundation (2022-3361), the Gordon and Betty Moore Foundation (GBMF7947), and the Alfred P. Sloan Foundation (G-2019-12502).    YT was supported by the Masason Foundation. 

\end{acknowledgments}

\newpage

\bibliography{biblio}

\newpage

\setcounter{equation}{0}
\setcounter{figure}{0}
\setcounter{table}{0}
\setcounter{section}{0}

\renewcommand{\thefigure}{S\arabic{figure}}
\renewcommand{\theequation}{S\arabic{equation}}
\renewcommand{\thetable}{S\arabic{table}}

\widetext
\begin{center}
\textbf{\large Supplementary Materials: Characterizing the Fundamental Bending Vibration of a Linear Polyatomic Molecule for Symmetry Violation Searches}
\end{center}
\section{Target Composition}

The data were obtained from targets of pressed Yb(OH)$_3$ powder in a stoichiometric mixture with Yb powder. The powders were mixed to have a 1:1 ratio of Yb and OH, ground using a mortar and pestle, passed through a 230 mesh sieve, and mixed with 4\% PEG8000 binder by weight. The powders were pressed in an 8~mm diameter die at a pressure of $\sim$1 GPa for $\sim30$ minutes. For some targets, 10-30\% water by mass was added to the powder before pressing, and while pressing the die was heated to $\sim$150\degree{}C. This was found to improve target density and ablation yield consistency.

\section{Phase Conventions} \label{sec:phase}

\subsection{$\Lambda$-Doubling}

There is an accepted convention for $\Lambda$-doubling, which was laid out by Mulliken and Christy \cite{Mulliken1931}. The convention is reiterated by Brown in \cite{Brown1979} and Brown and Carrington in \cite{Brown2003}. This convention is given by

\begin{equation} \label{eqn:Lambda_conv}
    \langle \Lambda = \pm 1 | e^{\pm2 i \phi_e} | \Lambda^\prime = \mp 1\rangle = -1 \times \delta_{\Lambda,\Lambda^\prime \pm 2}
\end{equation}
Here, $e^{\pm i \phi_e}$ is a raising/lowering operator with $\phi_e$ the azimuthal angle of the electrons. In this convention, a positive $q_e$ electronic $\Lambda$-doubling parameter in a ${}^1 \Pi$ state corresponds to the $(-1)^J$ parity level lying above the $(-1)^{J+1}$  parity level. In other words, the $+$ parity state is below the $-$ parity state for $J=1$. In the YbOH $\tilde{A}$ state, $p_e+2q_e$ is negative, and the $-$ parity state is below the $+$ parity state. This phase choice also manifests in the signs of the $\Lambda$-doubling Hamiltonian. When written in Hund's case (a), the $J_\pm S_\pm$ terms have a positive prefactor, and the $J_\pm J_\pm$ terms have a negative prefactor. For this work, we drop the $J_\pm J_\pm$ term in $\tilde{A}$ as its contribution is negligible. 

Now we derive the $\Lambda$ phase convention, following arguments from \cite{Hirota1985} and \cite{Brown2003}. We begin with the definition of the $L_z$ angular momentum operator in the molecule frame:

\begin{equation}
    L_z |\Lambda \rangle = - i \frac{\partial}{\partial \phi_e} | \Lambda \rangle = \Lambda |\Lambda \rangle
\end{equation}
This means $|\Lambda\rangle \propto e^{i \Lambda \phi_e}$. Since $L$ is not well defined, we expand $|\Lambda\rangle$ in terms of spherical harmonics:

\begin{equation}
    |\Lambda\rangle = \sum_L F_L Y_{L\Lambda}(\theta_e , \phi_e) = \sum_L \frac{F_L}{\sqrt{2\pi}} e^{i\Lambda \phi_e} \Theta_{L\Lambda}(\theta_e)
\end{equation}
Here, $\sum_L |F_L|^2 = 1$, and $\Theta_{L\Lambda}(\theta_e)$ is proportional to the associated Legendre functions $P^\Lambda_L(\cos{\theta_e})$. 

\begin{equation}
\begin{split}
    \Theta_{l, m} (\theta) &= (-1)^m \sqrt{\frac{2l+1}{2} \frac{(l - m)!}{(l+m)!}} P^m_l(\cos{\theta}) \text{\quad for } m\geq 0 \\
    & = (-1)^m \Theta_{l, -m}(\theta) \text{\quad for } m<0
    \end{split}
\end{equation}
Note the function $\Theta_{L\Lambda}$ satisfies $\Theta_{L,-|\Lambda|} = (-1)^\Lambda \Theta_{L,|\Lambda|}$. This is the origin of this specific phase-convention.

Now we can evaluate the left hand side of eqn. \ref{eqn:Lambda_conv}

\begin{equation}
\begin{split}
    \langle \Lambda | e^{\pm2 i \phi_e} | \Lambda^\prime\rangle &= \int \sin{\theta_e} \text{d}\theta_e \text{d}\phi_e \sum_{L,L^\prime} F^*_L F_{L^\prime} Y_{L\Lambda}(\theta_e,\phi_e)^* e^{\pm 2 i\phi_e} Y_{L^\prime \Lambda^\prime} (\theta_e,\phi_e)\\
        & = \sum_{L,L^\prime} F^*_L F_{L^\prime} \delta_{\Lambda, \Lambda^\prime\pm2} \int \sin{\theta_e} \text{d}\theta_e (-1)^{\Lambda^\prime \pm 2} \Theta_{L,-\Lambda^\prime \mp 2 }(\theta_e) \Theta_{L^\prime, \Lambda^\prime} (\theta_e)
\end{split}
\end{equation}

Where we substitute $Y_{L, \Lambda}(\theta_e, \phi_e)^* = (-1)^\Lambda Y_{L,-\Lambda}(\theta_e,\phi_e)$ and performed the $\phi_e$ integral taking advantage of the orthogonality of exponential functions. 

Now we simplify the integrand by noting we are interested in $\Lambda = \pm1, \Lambda^\prime = \mp 1$. This allows us to write $-\Lambda^\prime\mp2 = \Lambda^\prime$. Then the remaining $\theta_e$ integral can be performed by using the orthogonality relations of the associated Legendre polynomials, which results in 

\begin{equation}
\begin{split}
    \langle \Lambda = \pm 1 | e^{\pm2 i \phi_e} | \Lambda^\prime = \mp 1\rangle &=  \delta_{\Lambda, \Lambda^\prime\pm2}  \sum_L |F_L|^2 (-1)^{\Lambda^\prime} = -1 \times  \delta_{\Lambda, \Lambda^\prime\pm2} 
\end{split}
\end{equation}
Where we have substituted $|\Lambda| = 1$ in the last line and used the fact that $|F_L|^2$ is normalized. 

We also note that the behavior of $Y_{L \Lambda}$ upon the transformation $\Lambda \rightarrow -\Lambda$ gives the parity properties of $|\Lambda\rangle$. The action of space-fixed inversion, i.e. the parity operator $\mathcal{P}$, is equivalent to a reflection $\sigma_{xz}$ of the $xz$ plane of the molecule. This can be derived by considering the effect of space-fixed inversion on the Euler angles relating the molecule and lab frames. Therefore we have:

\begin{equation}
\raggedleft
    \begin{split}
        \mathcal{P} Y_{L, \Lambda} (\theta_e, \phi_e) &= \sigma_{xz} Y_{L,\Lambda} (\theta_e, \phi_e) \\
        &= Y_{L,\Lambda} (\theta_e, 2\pi - \phi_e) \\
        &= Y_{L,\Lambda} (\theta_e, \phi_e)^*  \\
        &= (-1)^\Lambda Y_{L,-\Lambda} (\theta_e, \phi_e)
    \end{split}
\end{equation}
This recovers the result $\mathcal{P} |\Lambda\rangle = (-1)^\Lambda |-\Lambda\rangle$ (note a $\Sigma^-$ state has an extra factor of $(-1)$ that we do not consider).  

For the full parity of the rotational wavefunction, the action of $\mathcal{P}$ must also be computed on the spin and rotational wavefunctions, which also reverse the projection quantum numbers and contribute parity phases of $(-1)^{S-\Sigma}$ and $(-1)^{J-\Omega}$ respectively. The combination of all phase factors gives the complete case (a) parity phase without bending motion: $(-1)^{\Lambda + S - \Sigma + J - \Omega} = (-1)^{J-S}$, where we have used $|\Sigma| = S$ and $\Omega = \Lambda + \Sigma$ to simplify the exponent. 

\subsection{$\ell$-Doubling}

For the derivation of the parity phase and matrix elements involving $\ell$, we follow Ref. \cite{Hirota1985}, which uses the vibrational phase conventions established by by Di Lauro and Mills \cite{DiLauro1966Coriolis}. The wavefunction for an isotropic 2D harmonic oscillator may be written as

\begin{equation}
    |v_2, \ell\rangle = \frac{1}{\sqrt{2\pi}} e^{i\ell \phi_n} \Psi_{v_2, \ell}(q)
\end{equation}
Here, $q=\sqrt{q_1^2 + q_2^2}$, where $(q_1, q_2)$ are the dimensionless, doubly-degenerate normal coordinates of the bending mode, and $\phi_n = \tan^{-1}(q_2/q_1)$ is the azimuthal angle of the bending nuclear framework. The function $\Psi_{v_2,\ell}$ is given by \cite{DiLauro1966Coriolis}:

\begin{equation}
    \Psi_{v,\ell} (q) = (-1)^{(v+|\ell|)/2} N_{v,\ell} q^{|\ell|} e^{-q^2/2} L^{|\ell|}_{(v+|\ell|)/2} (q^2) 
\end{equation}
Here, $N_{v,\ell}$ is a normalization factor and $L^k_n(x)$ is an associated Laguerre polynomial. This function explicitly  satisfies $\Psi_{v_2,|\ell|} = \Psi_{v_2,-|\ell|}$. 

We now consider the matrix elements between $\ell=\pm1$ states:

\begin{equation}
    \langle \ell | e^{\pm2 i \phi_n} | \ell^\prime\rangle = \int \text{d}q \text{d}\phi \frac{1}{2\pi} e^{-i\ell\phi_n} \Psi_{v,\ell}(q) e^{\pm 2 i \phi_n} e^{i\ell^\prime \phi_n} \Psi_{v,\ell^\prime} (q)
\end{equation}
The integration bounds are taken for $q \geq 0$ and $2\pi > q \geq 0$. The $\phi_n$ integral is evaluated with the orthogonality of complex exponential functions and enforces $\delta_{\ell, \ell^\prime+2}$. 

Restricting our attention to $\ell=\pm1$ states, the $\Psi_{v,\ell}(q)$ functions depend only on $|\ell|$, and do not add an additional phase. As a result we can evaluate the remaining $\text{d}q$ integral using the orthogonality relations of the associated Laguerre polynomials. We are left with

\begin{equation}
 \langle \ell | e^{\pm2 i \phi_n} | \ell^\prime\rangle = 1 \times \delta_{\ell,\ell^\prime\pm 2}   
\end{equation}
 
The difference between parity phase factors for $\ell$ and $\Lambda$ can be traced to the difference in phase between $\Psi_{v\ell}(q)$ and $\Theta_{L\Lambda}(\theta_e)$ upon space-fixed inversion. By considering the behavior of the wavefunctions under $\phi_n \rightarrow 2\pi - \phi_n$, we see the radial $q$ part is unaffected, giving us $\mathcal{P}|v_2, \ell\rangle = |v_2, -\ell\rangle$. When combined with rotational and spin parity phase factors, we then obtain the complete parity phase $(-1)^{J-S-\ell}$.

\section{Effective Hamiltonians and Matrix Elements} 

\subsection{\APi{}}

For the \APi{} state, we follow Ref. \cite{Steimle2019}, which uses the $R^2$ rotational Hamiltonian formalism. Details can be found in Ref. \cite{Brown2003}, Ch. 7. The effective Hamiltonian in Hund's case (a) and in spherical tensor notation is given by

\begin{equation}
\begin{split}
    H_{\tilde{A}} &=  T_0 + A  T^1_{q=0}(L) T^1_{q=0}(S) + B (J-L-S)^2 - D (J-L-S)^4  \\
    & + (p_e + 2q_e)\sum_{q=\pm1} e^{-2iq \theta} T^2_{2q}(J,S) + \frac{1}{2}(p_{eD} + 2q_{eD}) \sum_{q=\pm1} [N, e^{-2iq \theta} T^2_{2q}(J,S)]_+ \\
\end{split}
\end{equation}

Here, $T_0$ is the state origin, $A$ is the spin-orbit constant, $B$ is the rotational constant, $D$ is the centrifugal distortion term, $p_e+2q_e$ represents electronic $\Lambda$-doubling, $p_{eD}+2q_{eD}$ is the centrifugal distortion correction to $\Lambda$-doubling, $[\cdot,\cdot]_+$ is the anti-commutator, $J_\pm S_\pm$ are defined in the molecule frame, and $\theta$ is the azimuthal angle of the electronic wavefunction.  Matrix elements of this Hamiltonian can be found in \cite{Brown1978AlF, Hirota1985, Brown2003}. Note the $L_x^2 + L_y^2$ terms that arise in the $R^2$ formalism are absorbed in the origin. We use the constants determined in Ref. \cite{Steimle2019}. 

To be explicit, using the phase convention from supplementary section \ref{sec:phase} we reproduce our matrix element for the $\Lambda$-doubling term below:

\begin{equation}
    \begin{split}
        \langle \Lambda; S, \Sigma; J,\Omega, M| e^{2iq\theta} T^2_{2q}(J,S) |& \Lambda^\prime; S, \Sigma^\prime; J^\prime, \Omega^\prime, M^\prime \rangle \\
        & = \delta_{J,J^\prime} \delta_{M,M^\prime}  \delta_{\Lambda+2q,\Lambda^\prime} \\
        & \times (-1)^{J-\Omega} \threeJ{J}{-\Omega}{1}{-q}{J}{\Omega^\prime}  \sqrt{J (J+1)(2J+1)}  \\
        & \times (-1)^{S-\Sigma} \threeJ{S}{-\Sigma}{1}{q}{S}{\Sigma^\prime} \sqrt{S(S+1)(2S+1)} \\
    \end{split}
\end{equation}

\subsection{\XSigbend{}}

We reproduce the \XSigbend{} Hamiltonian below in spherical tensor notation. 

\begin{equation}
\begin{split}
    & H_{\tilde{X}} = T_0 + B (N^2-\ell^2) + \gamma \left(N\cdot S - T^1_{q=0}(N) T^1_{q=0}(S)\right) \\
    & + \gamma_G T^1_{q=0}(N) T^1_{q=0}(S) + \sum_{q=\pm1} e^{-2iq \phi} \left(p_G T^2_{2q}(N,S) - q_G T^2_{2q}(N,N)\right)
\end{split}
\end{equation}

\begin{figure}[t]
    \centering
    \includegraphics[width=0.25\textwidth]{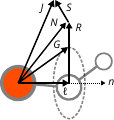}
    \caption{\label{fig:case_b} A schematic of the coupling scheme in Hund's case (b), used to describe the \Xbend{} state. The vibrational angular momentum $G$ is projected onto the internuclear axis to form $\ell$. The molecule rotation $R$ is coupled to $\ell$ to form $N$. Finally the spin-rotation interaction couples $S$ and $N$ to form $J$. Coupling to the H nuclear spin is not pictured.}
\end{figure}

The bending mode energy levels are well represented by Hund's case (b) eigenstates. A pictorial representation of the coupling scheme is given in Figure~\ref{fig:case_b}. 

As mentioned in the main text, the spin rotation interaction is modified to account for the bending motion. Here we provide further explanation. In the effective Hamiltonian approach, the spin-rotation parameter receives contributions from various orders of perturbation theory, $\gamma = \gamma^{(1)} + \gamma^{(2)}+ \cdots$~\cite{Brown2003}. The first order term $\gamma^{(1)}$ results from the magnetic interaction between the electron spin and the magnetic dipole moment of the rotating molecule. In heavy molecules, the first order term is small compared to the dominant second order contribution $\gamma^{(2)}$, arising from off-diagonal spin-orbit and rotational perturbations. For linear molecules with $N_z=0$, the spin-rotation term $N\cdot S$ implicitly only contains contributions from $N_x S_x$ and $N_y S_y$. However for a bending molecule, since $N_z \neq 0$, we explicitly subtract away $N_z S_z$.

Matrix elements for the $N^2$ and $N\cdot S$ terms can be found in Refs. \cite{Hirota1985, Brown2003}. Here we reproduce matrix elements for the terms specific to the bending mode. 

\begin{equation}
    \begin{split}
      \langle \ell; N, S, J, M| T^1_{q=0}(N) T^1_{q=0}(S)&| \ell^\prime; N^\prime, S, J^\prime, M^\prime \rangle \\
         &= \delta_{J,J^\prime} \delta_{N, N^\prime} \delta_{M,M^\prime}  \delta_{\ell,\ell^\prime} \times \ell\\
         &  \times (-1)^{J + N^\prime + S} \sixJ{N}{S}{J}{S}{N}{1}   \\
         &  \times (-1)^{N-\ell} \threeJ{N}{-\ell}{1}{0}{N}{\ell} (2N+1)\\
         &  \times \sqrt{S(S+1)(2S+1)}
    \end{split}
\end{equation}

\begin{equation}
    \begin{split} 
        \langle \ell; N, S, J, M| T^2_{2q} (N,S) e^{-2iq\phi} &| \ell^\prime; N^\prime, S, J^\prime, M^\prime \rangle \\
        & = \delta_{J,J^\prime} \delta_{N,N^\prime} \delta_{M,M^\prime} \delta_{\ell,\ell^\prime+2q}\\
        & \times (-1)^{J + N + S} \sqrt{\frac{5}{2}} \sixJ{N}{S}{J}{S}{N}{1}   \\
        & \times \sqrt{S(S+1)(2S+1)} \\
        & \times \sqrt{3} \sixJ{2}{1}{1}{N}{N}{N} \sqrt{N(N+1)(2N+1)} \\
        & \times (-1)^{N-\ell} \threeJ{N}{-\ell}{2}{2q}{N}{\ell} (2N+1)\\
        & \times \sqrt{S(S+1)(2S+1)} 
    \end{split}
\end{equation}

\begin{equation}
    \begin{split}
        \langle \ell; N, S, J, M| T^2_{2q} (N,N) e^{-2iq\phi} &| \ell^\prime; N^\prime, S, J^\prime, M^\prime \rangle \\
        & = \delta_{J,J^\prime} \delta_{N,N^\prime} \delta_{M,M^\prime} \delta_{\ell,\ell^\prime+2q}\\
        & \times (-1)^{J + N + S} \sixJ{N}{J}{S}{J}{N}{0}   \\
        & \times  \sqrt{5}  \sixJ{2}{2}{0}{N}{N}{N} \\
        & \times \frac{1}{2 \sqrt{6}} \sqrt{(2N-1)(2N)(2N+1)(2N+2)(2N+3)}\\
        & \times (-1)^{N-\ell} \threeJ{N}{-\ell}{2}{2q}{N}{\ell} (2N+1)\\
    \end{split}
\end{equation}

\subsection{Stark and Zeeman Matrix Elements}

For \Xbend{}, the Stark and Zeeman matrix elements are given in Hund's case (b). For the Stark matrix element, we only consider the contribution from the dipole component along the molecular $z$ axis. 

\begin{equation}
    \begin{split}
     \langle \ell; N, S, J, M| T^1_p(d) &| \ell^\prime; N^\prime, S, J^\prime, M^\prime \rangle \\
     & = (-1)^{J-M} \threeJ{J}{-M}{1}{p}{J^\prime}{-M} \\
     & \times (-1)^{J^\prime+N+S+1} \sqrt{(2J+1)(2J^\prime+1)} \sixJ{N^\prime}{J^\prime}{S}{J}{N}{1}\\
     &\times (-1)^{N-\ell} \sqrt{(2N+1)(2N^\prime+1)} \threeJ{N}{-\ell}{1}{0}{N^\prime}{\ell^\prime}
    \end{split}
\end{equation}

\begin{equation}
    \begin{split}
     \langle \ell; N, S, J, M| T^1_p(S) &| \ell^\prime; N^\prime, S, J^\prime, M^\prime \rangle \\
     & = \delta_{\ell,\ell^\prime} (-1)^{J-M} \threeJ{J}{-M}{1}{p}{J^\prime}{-M} \\
     & \times (-1)^{J+N+S+1} \sqrt{(2J+1)(2J^\prime+1)} \sixJ{S}{J^\prime}{N}{J}{S}{1}\\
     & \times \sqrt{S(S+1)(2S+1)}
    \end{split}
\end{equation}

\section{Intensity Borrowing and Transition Dipole Moments}

\subsection{Hund's Case (b) to Case (a) Change of Basis}

The eigenstates of \Xbend{} are best described by Hund's case (b) wavefunctions, while the eigenstates of \A{} are described by Hund's case (a) wavefunctions. To calculate transitions, we convert between the two cases using the following formula from Brown \cite{Brown1976}:

\begin{equation} \label{eq:a2b}
    |N,K,S,J,M\rangle = \sum_{\Sigma, P} (-1)^{N-S+P} \sqrt{2N+1} \threeJ{J}{P}{S}{-\Sigma}{N}{-K} |S,\Sigma ; J, P, M\rangle
\end{equation}
Here, $P=\Lambda +\Sigma + \ell$, and $K=\Lambda + \ell$. Note this form is equivalent to that given by Hirota in Ref. \cite{Hirota1985}.

\subsection{Transition Dipole Moment}

The transition dipole moment (TDM) matrix element is evaluated in Hund's case (a):

\begin{equation}\label{eq:TDM}
\begin{split}
    \langle \ell;\Lambda; S, \Sigma; J, P, M | T^1_p(d) &|  \ell^\prime;\Lambda^\prime; S, \Sigma^\prime; J^\prime, P^\prime, M^\prime\rangle\\
    & = \delta_{\Sigma,\Sigma^\prime} \delta_{\ell,\ell^\prime} \\
    & \times (-1)^{J-M} \threeJ{J}{-M}{1}{p}{J^\prime}{M^\prime}\\
    &\times \sqrt{(2J+1)(2J^\prime+1)} (-1)^{J-M} \\
    &\times \sum_q \threeJ{J}{-P}{1}{q}{J^\prime}{P^\prime} \delta_{\Lambda,\Lambda^\prime+q} \\
    & \times \langle \Lambda || T^1_q(d) || \Lambda^\prime\rangle
\end{split}
\end{equation}
The last term is the reduced matrix element encoding the transition dipole integral between two electronic states. The $\Delta\ell = 0$ selection rule is explicit in the above matrix element. This means we can only drive $\tilde{X}(010)$ to admixtures in $\tilde{A}(000)$ with $|\ell|=1$. 
 
\subsection{Mixing with $|\ell|=1$ states}

To model the transition intensities, as stated in the main text, we first separately diagonalize the \APi{} and \Xbend{} Hamiltonians. We then convert the $\tilde{X}(010)$ eigenvectors from Hund's case (b) to case (a), using equation \ref{eq:a2b}. 

Since we use effective Hamiltonians, the \APi{} eigenvectors have $\ell=0$. However, in reality these eigenvectors are perturbed by other states, and contain admixtures with $|\ell|=1$. These admixed states provide the transition intensity and non-zero transition dipole moment.

To represent the admixed states, we perform a change of basis to transform the \APi{} effective Hamiltonian eigenvectors into eigenvectors of the admixed states. As states in the main text, the states of interest with $\ell=1$ are $\tilde{A} \mu {}^2\Sigma^{(+)}_{1/2} (010)$, $\tilde{A} \kappa {}^2\Sigma^{(-)}_{1/2} (010)$, and $\tilde{B} {}^2 \Pi (010)$, where we are using vibronic term symbols $^{2S+1} K_P$. Each eigenvector of $\tilde{A}(000)$ is transformed into a linear combination of eigenvectors from the admixed states, with amplitudes $c_\mu,c_\kappa,c_B$. 

The mixing between \APi{} and $\tilde{B} ^2 \Pi (010)$ occurs at first order due to $H_{RT}$ (see main text). Since this interaction preserves $K$ and $P$, it simply exchanges one quanta between $\ell$ and $\Lambda$. Since \APi{} has $P=1/2$, we only consider mixing other $P=1/2$ states. We perform the following change of basis:

\begin{equation}
    \langle \tilde{B}(010), \Lambda=0,\ell,\Sigma, P \,{|}\,\tilde{A}(000), \Lambda^\prime,\ell^\prime = 0,\Sigma^\prime, P^\prime=\pm 1/2\rangle = \delta_{\ell, \Lambda^\prime} \delta_{P,P^\prime} 
    \delta_{\Sigma, \Sigma^\prime} (-1)^{P-1/2}
\end{equation} 
Note the phase factor $(-1)^{P-1/2}$ is explicitly included to preserve parity. This accounts for the extra $(-1)^\ell$ phase factor in the parity of an $\ell\neq0$ state compared to an $\ell = 0$ state. This factor can arise naturally if $H_{RT}$ is written as $\propto\sin{(\theta-\phi)}$ instead of being $\propto\cos{(\theta-\phi)}$. While the latter form is most often found in the literature~\cite{Brown2000, Hirota1985}, the former can be found in Ref.~\cite{Koppel1981} in the context of $\Sigma^-$ states.

The admixture of the $\mu$ and $\kappa$ states occurs via a second-order combination of $H_{RT}$ and $H_{SO}$. These interactions preserve $P$ but can change $K$. For $\mu(010)$ we obtain the following change of basis:

\begin{equation}
    \langle \mu(010), \Lambda,\ell,\Sigma, P \,{|}\, \tilde{A}(000), \Lambda^\prime,\ell^\prime = 0,\Sigma^\prime, P^\prime=\pm 1/2\rangle = \delta_{\Lambda,-\Lambda^\prime} \delta_{\ell,\Lambda^\prime} \delta_{\Sigma,-\Sigma^\prime} (-1)^{P-1/2}
\end{equation} 
And for $\kappa(010)$:

\begin{equation}
    \langle \kappa(010), \Lambda,\ell,\Sigma, P | \tilde{A}(000), \Lambda^\prime,\ell^\prime = 0,\Sigma^\prime, P^\prime=\pm 1/2\rangle = \delta_{\Lambda,\Lambda^\prime} \delta_{\ell,-\Lambda^\prime} \delta_{\Sigma,-\Sigma^\prime} (-1)^{P-1/2}
\end{equation} 

After changing basis to states with $|\ell|=1$, we compute the transition dipole matrix element using equation \ref{eq:TDM}. The transition amplitudes for the different state admixtures are added together, and the resulting interference depends on the mixing coefficients $c_\mu,c_\kappa, c_B$. Finally, to obtain relative intensities, we square the total transition amplitude.

\section{Assigned Lines}

\begingroup
\begin{table*}[ht]
\begin{threeparttable}
\caption{\label{tab:lines}
Observed lines, ground states quantum numbers ($N^{\prime\prime}, J^{\prime\prime}, \mathcal{P}^{\prime\prime}$), excited states quantum numbers ($J^\prime, \mathcal{P}^\prime$), observed positions, and residuals of $\tilde{X}{}^2\Sigma^+(010)\rightarrow\tilde{A}{}^2\Pi_{1/2}(000)$ band of YbOH. There are in total 38 lines assigned to 39 transitions as the ${}^Q R_{12}^{-} (1)$ and ${}^P Q_{12}^{-}(5)$ lines are overlapped. The fit residual is 6.1 MHz.}
\begin{ruledtabular}
\begin{tabularx}{\textwidth}{lllld}
Line	&	$N^{\prime\prime}, J^{\prime\prime}, \mathcal{P}^{\prime\prime}$	&	$J^\prime, \mathcal{P}^{\prime}$	&	Obs. (cm$^{-1}$)	&	\multicolumn{1}{c}{Obs. - Calc. (MHz)} \\ \hline \\[-0.8em]

 ${}^{O} {P}_{1{2}}^+$ & 2, 3/2, + & 1/2, $-$ & 17002.4883 &  4.4 \\
  & 3, 5/2, + & 3/2, $-$ & 17002.4312 & -7.4 \\
  & 4, 7/2, + & 5/2, $-$ & 17000.6512 & -2.7 \\

\\
 ${}^{O} {P}_{1{2}}^-$ & 2, 3/2, $-$ & 1/2, + & 17002.9232 & -0.1 \\
 & 3, 5/2, $-$ & 3/2, + & 17001.5614 & 14.9 \\
\\
 ${}^{P} {P}_{1{1}}^+$ & 1, 3/2, +  & 1/2, $-$ & 17003.4683 & -0.2 \\
   & 3, 7/2, +  & 5/2, $-$ & 17002.6114 &  1.7 \\
  & 5, 11/2, + & 9/2, $-$ & 17001.8212 & 12.2 \\
\\
 ${}^{P} {P}_{1{1}}^-$ & 1, 3/2, $-$ & 1/2, + & 17003.9070 & -2.2 \\
  & 2, 5/2, $-$ & 3/2, + & 17003.0314 & -3.6 \\
  & 4, 9/2, $-$ & 7/2, + & 17002.2076 & -4.8 \\
\\
 ${}^{P} {Q}_{1{2}}^+$ & 2, 3/2, + & 3/2, $-$ & 17003.9039 & -8.8 \\
  & 3, 5/2, + & 5/2, $-$ & 17002.6012 & -5.8 \\
  & 5, 9/2, + & 9/2, $-$ & 17001.8046 & 12.7 \\
\\
 ${}^{P} {Q}_{1{2}}^-$ & 1, 1/2, $-$ & 1/2, + & 17003.9053 & -5.8 \\
  & 2, 3/2, $-$ & 3/2, + & 17003.0250 & -5.0 \\
  & 3, 5/2, $-$ & 5/2, + & 17003.9208 & -5.3 \\
  & 5, 9/2, $-$ & 9/2, + & 17004.0076 & 13.3 \\
\\
 ${}^{Q} {Q}_{1{1}}^+$ & 1, 3/2, +  & 3/2, $-$  & 17004.8846 &   3.3 \\
  & 3, 7/2, +  & 7/2, $-$  & 17005.9150 & -13.0 \\
  & 5, 11/2, + & 11/2, $-$ & 17007.0123 &  -3.5 \\
 \\
 ${}^{Q} {Q}_{1{1}}^-$ & 1, 3/2, $-$  & 3/2, +  & 17004.0091 &   5.5 \\
  & 2, 5/2, $-$  & 5/2, +  & 17005.3917 &  -0.7 \\
  & 4, 9/2, $-$  & 9/2, +  & 17006.4556 &  -1.6 \\
\\
 ${}^{Q} {R}_{1{2}}^+$ & 1, 1/2, + & 3/2, $-$ & 17004.8824 &  1.3 \\
  & 2, 3/2, + & 5/2, $-$ & 17004.0743 &  5.1 \\
  & 3, 5/2, + & 7/2, $-$ & 17005.9052 & -5.8 \\
\\
 ${}^{Q} {R}_{1{2}}^-$ & 1, 1/2, $-$ & 3/2, + & 17004.0076 &  7.3 \\
  & 2, 3/2, $-$ & 5/2, + & 17005.3853 &  0.9 \\
  & 4, 7/2, $-$ & 9/2, + & 17006.4421 & -6.9 \\
 \\
 ${}^{R} {R}_{1{1}}^+$ & 1, 3/2, +  & 5/2, $-$  & 17005.0543 & -1.5 \\
  & 2, 5/2, +  & 7/2, $-$  & 17007.3837 & -0.7 \\
  & 3, 7/2, +  & 9/2, $-$  & 17006.2215 &  2.3 \\
  & 4, 9/2, +  & 11/2, $-$ & 17009.4646 & -2.9 \\
 \\
 ${}^{R} {R}_{1{1}}^-$ & 1, 3/2, $-$  & 5/2, +  & 17006.3695 & 12.7 \\
  & 2, 5/2, $-$  & 7/2, +  & 17005.6298 &  6.1 \\
  & 3, 7/2, $-$  & 9/2, +  & 17008.4157 &  3.7 \\
  & 4, 9/2, $-$  & 11/2, + & 17006.8298 & -0.6 \\
  & 5, 11/2, $-$ & 13/2, + & 17010.5312 & -0.9 \\

\end{tabularx}
\end{ruledtabular}
\end{threeparttable}
\end{table*}
\endgroup

See Table \ref{tab:lines}. Line notation is described in the main text. 

\end{document}